\documentclass[%
 reprint,
 amsmath,amssymb,
 aps,
]{revtex4-2}
\usepackage{comment}
\usepackage{placeins}
\usepackage{graphicx}
\usepackage{gensymb}
\usepackage{dcolumn}
\usepackage{bm}

\usepackage[justification=Justified,singlelinecheck=false]{caption}
\usepackage{amsmath}
\usepackage{booktabs}
\usepackage{svg}

\usepackage{physics}

\newcommand{\AlO}{\ensuremath{\mathrm{Al}_2\mathrm{O}_3}}
\newcommand{\SiN}{\ensuremath{\mathrm{Si}_3\mathrm{N}_4}}
\newcommand{\SiO}{\ensuremath{\mathrm{Si}\mathrm{O}_2}}

\newcommand{\Ca}{\ensuremath{^{40}\mathrm{Ca}^+}}
\newcommand{\sigmap}{\hat{\bm{\sigma}}_+}
\newcommand{\sigmam}{\hat{\bm{\sigma}}_-}
\newcommand{\Sm}{|4S_{1/2}, m_j = -1/2\rangle}
\newcommand{\Sp}{|4S_{1/2}, m_j = +1/2\rangle}
\newcommand{\Pm}{|4P_{1/2}, m_j = -1/2\rangle}
\newcommand{\Pp}{|4P_{1/2}, m_j = +1/2\rangle}
\newcommand{\x}{\hat{\bm{x}}}
\newcommand{\y}{\hat{\bm{y}}}
\newcommand{\z}{\hat{\bm{z}}}
\newcommand{\nbar}{\bar{n}}
\newcommand{\tpi}{2 \pi \times}

\begin{document}
\preprint{APS/123-QED}

\title{Rapid multi-mode trapped-ion laser cooling in a phase-stable standing wave}

\author{Zhenzhong Xing$^{\S,1}$}
 \email{zx296@cornell.edu} 
 \author{Hamim Mahmud Rivy$^{\S, 1}$}
 \email{hr296@cornell.edu \\ $^\S$ These authors contributed equally}

\author{Vighnesh Natarajan$^{1}$}
\author{Aditya Milind Kolhatkar$^{1}$}
\author{Gillenhaal Beck$^{2}$}
 \author{Karan K. Mehta$^{1}$}
\email{karanmehta@cornell.edu}
\affiliation{$^{1}$School of Electrical and Computer Engineering, Cornell University, Ithaca, NY 14853}
\affiliation{$^{2}$Institute for Quantum Electronics, ETH Zurich, 8093 Zurich, Switzerland}

\date{\today}

\begin{abstract}
Laser cooling is fundamental to quantum computing and metrology using atomic systems. Precise control often requires cooling atoms' motional degrees of freedom to the quantum ground state, imposing operation time and architectural limitations particularly in large-scale systems. Here we demonstrate how the integrated optical control of interest for scaling trapped-ion systems additionally enables laser cooling that bypasses limitations of conventional schemes. Leveraging multi-channel integrated delivery of ultraviolet to infrared wavelengths for calcium ion control including in passively phase-stable ultraviolet standing waves (SWs),  we experimentally verify a long-standing prediction by Cirac et al., realizing Doppler cooling to below the conventional Doppler limit at a SW node. We also present the first realization of ground-state cooling via electromagnetically induced transparency (EIT) using a ``probe" beam delivered as a SW with atoms positioned at a node, predicted to enable multi-mode sub-recoil-limit laser cooling. We demonstrate cooling of motional modes spanning an approximately 5 MHz bandwidth from the Doppler temperature to near the ground state within 150~\textmu s, reaching $\bar n \approx 0.05$ phonon number occupancies for the target mode. Direct evaluation against the comparable running-wave (RW) scheme shows the SW implementation's simultaneous advantage in cooling rate, motional mode bandwidth, and final phonon number, as previously theoretically predicted. Our work leverages capabilities enabled by integrated optical delivery to demonstrate fast cooling of multiple modes to the quantum ground state, and more broadly how scalable approaches to optical control can enable enhancements in fundamental atomic functionalities.
\end{abstract}

\maketitle

\section{Introduction}
Trapped-ion quantum computing (QC) \cite{bruzewicz2019trapped} and metrology rely on laser cooling for precise coherent control. State-of-the-art QC systems based for example on quantum charge-coupled device (QCCD) schemes \cite{kielpinski2002architecture, home2009complete, pino2021demonstration} employ repeated ground-state laser cooling to mitigate unintended excitation following transport and ion crystal reconfiguration operations \cite{bowler2012coherent, van2020coherent, burton2023transport} and to mitigate inevitable heating during operation \cite{brownnutt2015ion}. In many QC architectures, total runtime can be dominated by transport and repeated ground-state cooling \cite{pino2021demonstration, ransford2025helios}, often implemented via resolved sideband cooling of individual modes \cite{monroe1995resolved}. Cooling times required can be severe given that high-fidelity laser-based quantum logic often requires low phonon number occupancies for multiple motional modes spread over multiple MHz bandwidths \cite{ballance2017high}. Polarization-gradient \cite{ejtemaee20173d, joshi2020polarization} and electromagnetically induced transparency (EIT) cooling \cite{roos2000experimental, lechner2016electromagnetically} offer routes to sub-Doppler cooling for multiple modes simultaneously, but are often limited to higher phonon numbers than those achievable with sideband cooling. 

Optical delivery leveraging waveguide photonics integrated in ion traps \cite{mehta2016integrated, niffenegger2020integrated, mehta2020integrated, ivory2021integrated} stands to address challenges related to scaling and parallelizing trapped-ion optical control \cite{moody20222022, kwon2024multi, mordini2025multizone}. Beyond scaling, integrated addressing brings passive phase and amplitude stability which enables practical delivery of spatially structured driving fields, and thereby tailoring of laser-ion interactions in ways that can alleviate limitations on basic physical operations. This has been proposed and explored for example via delivery of phase-stable standing waves (SWs) \cite{mehta2019towards, vasquez2023control, saner2023breaking, kolhatkar2025efficient} and polarization gradients \cite{clements2024sub, corsetti2024integrated}. 

Previous work has examined laser cooling in SWs theoretically, both for Doppler cooling \cite{cirac1992laser} and for schemes capable of reaching the quantum ground state based on electromagnetically induced transparency (EIT) \cite{roos2000experimental, lechner2016electromagnetically}. Due to the suppression of coherent carrier coupling for atoms located at SW nulls \cite{leibfried2003quantum} and the associated photon scattering undesired for laser cooling, theoretical treatments have predicted modest advantages in cooling speed and phonon number limits for Doppler cooling. For ground-state EIT cooling, use of a SW ``probe" or ``cooling" beam with ions positioned at intensity nulls has been predicted to offer significant improvements simultaneously in cooling rate, bandwidth, and final phonon number \cite{zhang2012dark, xing2025trapped}, but has not previously been experimentally realized. Importantly, due to the SW EIT scheme's complete suppression of both first-order carrier and blue sideband transitions in favor of the red sideband transitions that extract energy from motion, it in principle allows motional mode occupancies below the photon recoil limit and comparable to or below those achievable with resolved sideband cooling, furthermore for multiple modes simultaneously and at significantly higher rates.

In this work, we experimentally demonstrate fast, broadband laser cooling in nodal positions of integrated SW drive fields. We measure Doppler cooling to below the Doppler limit for a trapped ion positioned at SW nodes, realizing a long-standing prediction and verifying the basic mechanism. We further experimentally demonstrate SW EIT ground-state cooling of radial motional modes to phonon occupancies $\bar n \approx 0.05$, and all single-ion motional modes spanning a 5 MHz bandwidth close to the ground state ($\bar n \lesssim 1$), within 150 \textmu s starting from near the Doppler limit. The foundry-fabricated trap devices utilized in this work incorporate running-wave (RW) beam delivery that allows direct experimental comparison between the carrier-nulled SW scheme and the RW equivalent. We thereby demonstrate SW EIT's simultaneous advantages in motional mode bandwidth, cooling rate, and final phonon number. We present detailed simulations of limiting nonidealities in our apparatus that indicate straightforward design optimizations would reduce the final phonon number limit by over an order of magnitude further, to the limits imposed by current technical noise levels. This work realizes fast, multi-mode ground-state cooling, addresses a significant operation time limitation in trapped-ion QC systems, and demonstrates a clear performance advantage practically enabled by scalable optical control of atomic systems.

In the following, we first review the basic mechanisms and intuition behind the cooling schemes demonstrated here. We then describe the experimental platform and apparatus utilized and present characterization of the photonics spanning ultraviolet (UV) to near infrared (NIR) wavelengths employed in the trap device, including profiling of the UV SW using a single $\Ca$ ion as a high-resolution probe. We thereby characterize the extinction realized at the SW nodes, along with our ability to precisely position the atom in this spatially varying intensity profile. We present results demonstrating Doppler cooling to below the Doppler limit at SW nodes for multiple modes, followed by SW EIT cooling of all single-ion modes to near the ground state. We perform direct evaluation against the comparable RW EIT scheme to verify the SW field's role and the relative performance of the two schemes. Experimental details and numerical simulations quantifying technical limitations on the observed performance are presented in the Appendix, indicating concrete routes to further improvement. We conclude with a discussion of these straightforward improvements and the expected resulting performance, the fundamental limits of the cooling scheme demonstrated here, and extensions to multi-ion crystals.

\section{Carrier-nulled cooling mechanism}
\begin{figure}[t!]
    \centering
\includegraphics[width=1\columnwidth]{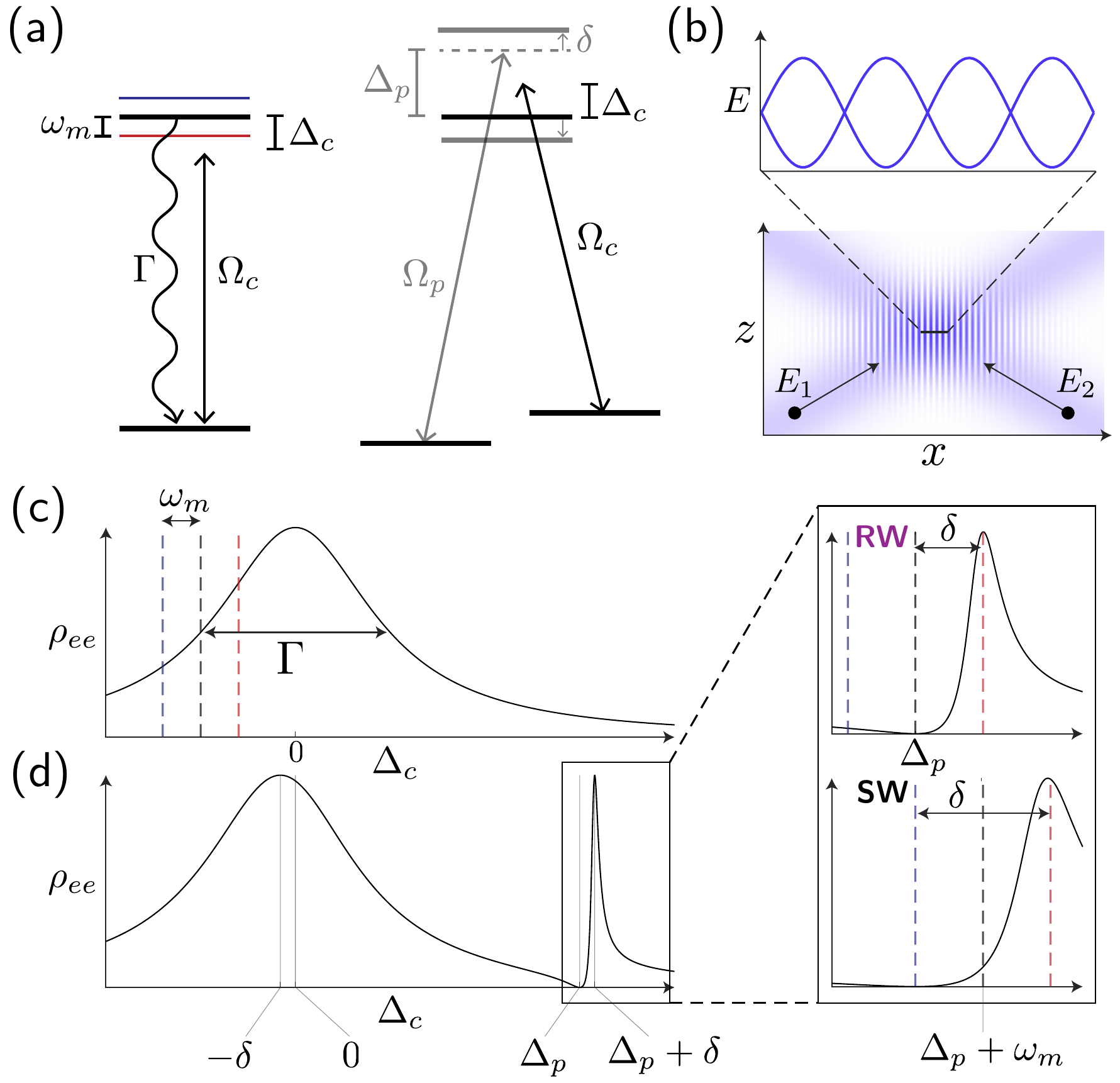}
    \caption{General mechanism behind SW cooling schemes demonstrated. (a) Basic level structures relevant to Doppler (left) and EIT cooling (right), with symbols as defined in the main text. Blue and red lines (shown only on the left for simplicity) indicate motional sideband transitions that raise or lower $n$ of a mode of frequency $\omega_m$. (b) Phase-stable beams counterpropagating along $x$ and polarized along $y$ in this illustration result in a standing-wave intensity profile; atoms confined at nodes experience maximum gradient that drives sideband transitions only, with nulled carrier excitation. (c) Excitation spectrum of the $\Omega_c$ beam for Doppler cooling. When $\Delta_c$ is chosen at $-\Gamma/2$; probabilities of RSB and BSB scattering are proportional to $\rho_{ee}$ at the red and blue dashed lines, and similarly for carrier scattering in a RW at the black dashed line. (d) $\Omega_c$ excitation spectrum for EIT cooling in the presence of $\Omega_p$ at fixed $\Delta_p$. Insets show choice of pump beam strength $\Omega_p$ and resultant AC stark shift $\delta_\mathrm{EIT}$ for optimal cooling of a mode of motional frequency $\omega_m$, for the $\Omega_c$ beam delivered as a RW, or as a SW. 
    \label{fig:SWmech}}
\end{figure}

\begin{figure*}[ht!]
\centering
\includegraphics[width=1\textwidth]{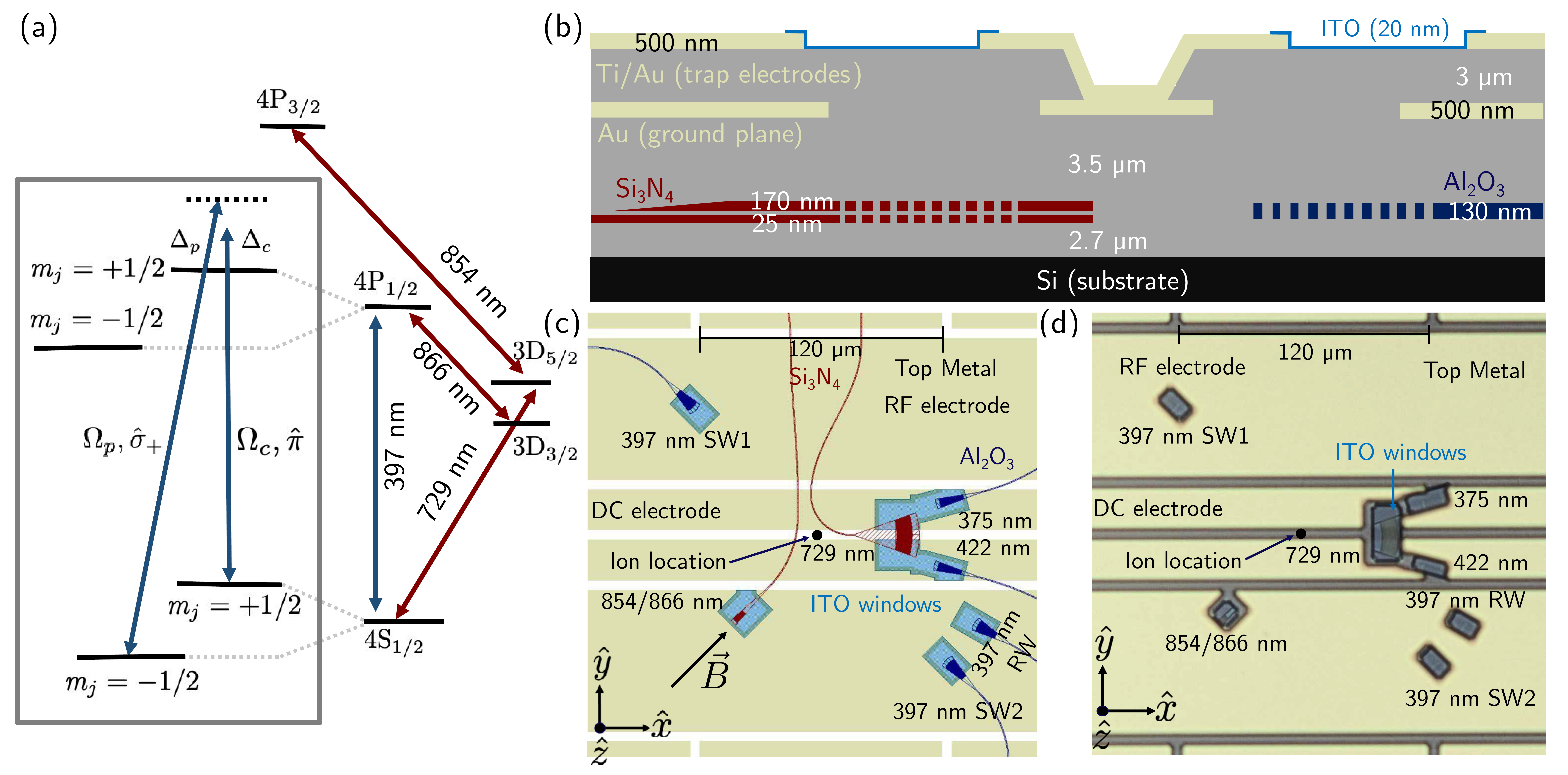}
\caption{Experimental platform and ion trap device delivering the required wavelengths. (a) Relevant energy level structure of \Ca{}. $\lambda=397$~nm transitions are used for laser cooling and state readout; light at 729 nm drives an electric quadrupole transition for spectroscopy and motional diagnostics; and 854 and 866 nm serve as repumpers. $\Delta_p$ and $\Delta_c$ denote the detuning of the EIT pump ($\hat{\bm\sigma}_+$) and the probe beam ($\hat{\bm \pi}$) from the $\ket{4S_{1/2},m_j={-1/2}} \leftrightarrow \ket{4P_{1/2},m_j={+1/2}}$ and the $\ket{4S_{1/2},m_j={+1/2}} \leftrightarrow \ket{4P_{1/2},m_j={+1/2}}$ transitions respectively.  (b) Layer stack-up schematic, incorporating waveguide features for routing and emission in a 130 nm-thick \AlO{} layer and in \SiN{} layers of 25 and 170 nm thickness. A 500-nm-thick Au layer (using Ti for adhesion) forms both the ground plane and the top RF and DC electrodes. A 20-nm-thick conductive ITO film is used to mitigate potential surface charging at top electrode openings. (c) Layout of the trap zone used, and the integrated waveguide and grating elements at the indicated wavelengths, $B$-field orientation, top metal, and ITO features. Additionally, free-space $\lambda=397$~nm beams are employed for the $\hat{\bm\sigma}_+$ EIT pump and for the $\hat{\bm\pi}$-polarized RW Doppler cooling beam, propagating parallel to the trap surface along and perpendicular to $\vec{B}$, respectively.  (d) Bright-field microscope image of the same zone.}
\label{fig:Ion trap device}
\end{figure*}

The enhancements to both Doppler cooling and EIT ground-state cooling demonstrated here rely on a similar basic mechanism \cite{cirac1992laser, zhang2012dark, xing2025trapped} which we briefly review in this section. Fig.~\ref{fig:SWmech}a shows the level structure for a closed two-level system of frequency $\omega_0$, driven by a cooling laser field with amplitude $\Omega_c$. Considering one mode of motion at frequency $\omega_m$ of such an atom in a harmonic trap, in addition to the ``carrier" transition, resonant at drive frequency $\omega_0$, resonant transitions at $\omega_0 \pm \omega_m$ add or substract a quantum of motion upon internal state excitation, referred to as ``blue" and ``red" sideband transitions (RSB and BSB, respectively) and indicated by the blue and red lines in Fig.\ref{fig:SWmech}a. For atoms with wavefunction extent small compared to the wavelength (the Lamb-Dicke regime) \cite{leibfried2003experimental} where first-order sidebands dominate, the carrier transition and these sideband transitions approximate the full behavior. For a two-level system with electric dipole moment, the carrier transition has amplitude $\Omega_c$ proportional to the electric field strength at the ion location. The sideband transitions have amplitude proportional to the field gradient, $\eta \Omega_c$, with $\eta$ the Lamb-Dicke parameter, well below unity for typical parameters. Hence for light delivered as a phase-stable standing wave (Fig.~\ref{fig:SWmech}b) to an atom positioned at a node where only the field gradient is nonzero, the carrier is nulled and only sidebands are driven. 

\begin{figure*}[ht!]
    \centering
\includegraphics[width=1\textwidth]{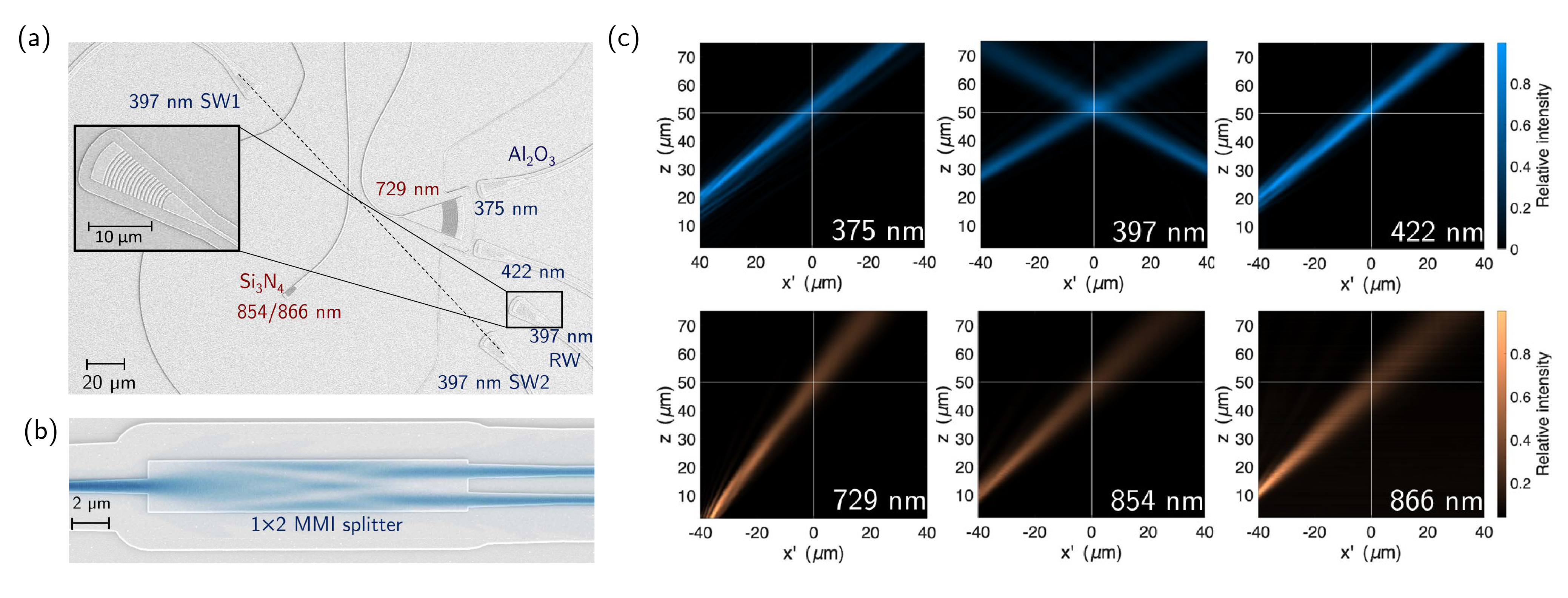}
    \caption{Integrated optical component characterization. Scanning electron microscope (SEM) images of (a) fabricated \AlO{} and \SiN{} waveguides features near the trap zone employed in this work, following waveguide fabrication and prior to cladding and metallization. (b) The $1 \times 2$ \AlO{} MMI splitter, overlaid with the simulated intensity profile for $\lambda=397$ nm, feeding the `SW1' and `SW2' grating couplers. (c) Measured grating emission profiles at the labeled wavelengths along the ``longitudinal" direction $x'$ for each grating (e.g., dashed line in (a) for the SW couplers), obtained from images of the radiated fields of the integrated grating couplers at different heights above the trap electrode; line crossings indicate the target ion position, demonstrating realized targeting accuracy of $\lesssim 2$ \textmu m for all wavelengths. }
    \label{fig:levstruc_wavelengths}
\end{figure*}

Cooling is generally achieved via excitation in a manner that preferentially drives RSB transitions. Fig.~\ref{fig:SWmech}c shows the excited-state population and hence spontaneous scattering rate as a function of a cooling beam detuning for an atom at rest. A RW cooling beam at $\Delta_c = -\Gamma/2$ results in optimal contrast between RSB and BSB scattering rates; however the strong carrier transition is also driven, resulting in additional photon scattering that heats the ion motion. When $\Omega_c$ is instead delivered as a SW and the ion positioned at a node, this carrier scattering is nulled, resulting in an approximately $2\times$ lower cooling limit \cite{cirac1992laser}, quantified by the steady-state expectation phonon number $\bar n_\mathrm{ss}$. Additionally, carrier suppression reduces power broadening of the line at high $\Omega_c$ which in turn would reduce contrast in the RSB and BSB scattering rates; this reduced saturation allows SW cooling to operate at higher $\Omega_c$ and hence cooling rates both within and beyond the LD regime \cite{cirac1992laser, xing2025trapped}.

A similar concept can be applied to EIT cooling, with more significant improvement in the cooling performance. EIT cooling utilizes a three-level electronic system driven by a strong pump beam ($\Omega_p, \Delta_p$) (Fig.~\ref{fig:SWmech}a) which modifies the atomic response to a temporally coherent ``cooling" or ``probe" beam ($\Omega_c$, $\Delta_c$). Excitation is suppressed at the two-photon resonance $\Delta_c = \Delta_p$, with bright resonances AC Stark shifted by $\delta_\mathrm{EIT} = \frac{1}{2} (\sqrt{\Omega_p^2 + \Delta_p^2}-\Delta_p)$ from either the single- ($\Delta_c=0$) or two-photon cooling beam resonance conditions (gray lines) \cite{scully1997quantum}. As shown in Fig.~\ref{fig:SWmech}d, RW EIT uses the dark resonance to suppress carrier excitation by setting $\Delta_c = \Delta_p$; by choosing $\Omega_p$ and $\Delta_p$ to set $\delta_\mathrm{EIT} = \omega_m$, the narrow bright resonance results in strong contrast between the RSB and BSB scattering rates, and much lower $\bar n_\mathrm{ss}$ limits as compared to Doppler cooling \cite{roos2000experimental, lechner2016electromagnetically}. However the finite BSB excitation rate still limits $\bar n_\mathrm{ss}$ values achievable, and the strong carrier coupling on the $\Omega_c$ beam again limits cooling rates. Delivery of $\Omega_c$ as a SW and setting $\delta_\mathrm{EIT} = 2\omega_m$ allows for a different choice of $\Delta_c$ (Fig.~\ref{fig:SWmech}d) that leverages the \emph{spatial} coherence of the interfering fields to null carrier excitation, and uses the EIT dark resonance to perfectly null BSB excitation. This results simultaneously in dramatically lower $\bar n_\mathrm{ss}$ values achievable, comparable to or even below resolved sideband cooling; significantly broader cooling bandwidth, i.e. the range of motional mode frequencies around the target $\omega_m$ that can be effectively cooled; along with higher cooling rates \cite{xing2025trapped}. 

\section{Experimental system}

The relevant energy levels and transitions of the \Ca{} ions used in this work are shown in Fig.~\ref{fig:Ion trap device}a. We use a foundry-fabricated ion trap chip device with both silicon nitride (\SiN{}) \cite{worhoff2015triplex} and aluminum oxide (\AlO{}) waveguides \cite{west2019low,sorace2019versatile, garcia2024uv} to deliver all required wavelengths for $\Ca$ ion control. The device layer stack-up is shown in Fig.~\ref{fig:Ion trap device}b. Devices are fabricated on 100 mm-diameter Si substrates, with thin-film \SiN{} and \AlO{} patterned via electron-beam lithography to define photonic elements for coupling, splitting, routing, and beam-forming. In this work, \AlO{} is used for the blue and UV wavelengths ($\lambda=375$, 397, and 422 nm) and \SiN{} for visible and near-IR wavelengths ($\lambda=729$, 854, and 866 nm). The latter are of the same cross section as employed in previous work \cite{mehta2020integrated}. The top Ti/Au layer forms a surface-electrode Paul trap \cite{chiaverini2005surface} designed to confine ions 50 \textmu m above the top surface. The openings for beam delivery from waveguides to the ion are coated with a 20 nm thick conductive indium-tin-oxide (ITO) layer \cite{niffenegger2020integrated}, intended to shield the ion from stray charge on exposed dielectrics, including light-induced charging. Underneath the trap electrodes, a gold ground plane is formed to shield the silicon substrate from RF fields and the ions from the mobile charge carriers in the substrate \cite{niedermayr2014cryogenic, mehta2014ion}.   

Figure~\ref{fig:Ion trap device}c shows a schematic of the trap zone and beam configuration used in this work, with waveguide channels and couplers for wavelengths used for photoionization  ($\lambda=375$ and 422 nm), optical transition coherent control (729 nm), repumping (854 and 866 nm), and state preparation, laser cooling, and readout (397 nm). Two grating couplers (`SW1' and `SW2') fed by the same $1 \times 2$ multi-mode interferometer (MMI) splitter emit at $\theta = 60^\circ$ from the trap normal to intersect 50 \textmu m above the trap surface, forming a phase-stable $\lambda=397$~nm SW in the $x-y$ plane. These are oriented along a line at $-45^\circ$ to the trap axis $\hat{\bm{x}}$ to produce wave vector components along the axial and both radial motional modes, and a $\hat{\bm\pi}$-polarized field when fed with quasi-TE waveguide modes given the orientation of the quantizing magnetic field. A separate coupler oriented $-34^\circ$ to the trap axis delivers the RW EIT probe beam, also approximately $\hat{\bm \pi}$-polarized. For EIT cooling, these $\bm{\hat\pi}$-polarized fields couple $\Sp$ and $\Pp$, while the $\hat{\bm\sigma}_+$-polarized beam serving as the EIT ``pump" coupling $\Sm$ and $\Pp$ is delivered via free-space optics and propagates along the magnetic field direction. An optical microscope image of the fabricated trap zone is shown in Fig.~\ref{fig:Ion trap device}d, with scanning electron microscope images of the waveguide features shown in Fig.~\ref{fig:levstruc_wavelengths}a and b, respectively. 

Fig.~\ref{fig:levstruc_wavelengths}c shows measured longitudinal emission profiles of the gratings at the designed wavelengths, profiled using a scanning imaging system with a scientific CCD camera \cite{mehta2017precise}. We measure a targeting inaccuracy of less than 2~\textmu m for all wavelengths, indicating the functioning of the optical components in accordance with design and simulation. This allows the experiments presented below to efficiently utilize delivery of the other required wavelengths via fiber-coupled waveguide paths alongside the $\lambda=397$~nm light most directly implicated in the cooling schemes studied. 

For cryogenic ion trap experiments, light is delivered to the integrated waveguides via a 14-channel fiber V-groove array (VGA) populated with single-mode (SM) fibers for both UV and visible/NIR wavelengths. The VGA and trap chip are aligned to each other and mutually attached to a common carrier PCB in a fashion compatible with both cooling and thermal cycling to the 4.8 K experimental operation temperature, as well as UV delivery without photodamage. In this work, all $\lambda=729$, $854$, and $866$~nm light is delivered through the integrated channels. The integrated SW and RW $\lambda=397$~nm paths are used for SW and RW cooling beam delivery for EIT experiments. Photoionization to load the trap utilizes the integrated $\lambda=422$~nm path and a free-space $\lambda=375$~nm beam. A free-space $\bm{\hat \sigma}_+$-polarized  $\lambda=397$~nm beam is used for state preparation and as the EIT pump beam, and a free-space $\bm{\hat \pi}$-polarized beam for RW Doppler cooling and readout. Further details on the full device layout and packaging are provided in the Appendix.

\section{SW profiling and characterization}

We use a single \Ca{} ion as a high-resolution field probe, and as a means to evaluate positioning stability in the UV SW generated via the on-chip optics. We control the ion position via DC voltages applied to the trap electrodes and characterize the SW profile by measuring the AC Stark shift $\delta$ on a narrow-linewidth optical quadrupole transition as a function of ion position along the trap axis (controlled via DC voltages applied to the trap electrodes), with $\lambda=397$~nm light detuned by $\Delta_{397}$ from the $\ket{4S_{1/2}} \leftrightarrow \ket{4P_{1/2}}$ transition incident through the SW couplers. We use a small detuning $\Delta_{397}/2\pi = -1$~GHz such that $\delta$ is dominated by the AC stark shift of $\Sp$. 

\begin{figure}[t!]
    \centering
\includegraphics[width=0.45\textwidth]{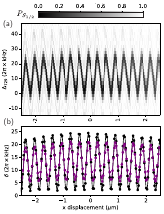}
    \caption{SW profiling via AC Stark shift measurements on the optical quadrupole transition. (a) Population $P_{S_{1/2}}$ remaining in $\ket{4S_{1/2}}$ after a $\pi$ pulse on the $|4S_{1/2}, m_j = +1/2\rangle \leftrightarrow |3D_{5/2}, m_j = +1/2\rangle$ carrier transition, with $\lambda = 397$~nm light detuned by $\Delta_{397}/2\pi = -1$~GHz sent to the SW couplers. The measurement is taken for varying axial displacements of the ion, with axial mode initially cooled to the ground state via RW EIT cooling for this data. $\Delta_{729} = 0$ corresponds to the quadrupole transition frequency calibrated with no light in the SW. (b) Corresponding fit AC stark shift on $\Sp$ $\delta$ vs. axial displacement (black), along with that measured only after Doppler cooling (purple). Lines correspond to a fit SW profile with Gaussian amplitude envelope. The contrast near the SW center at 0 displacement for ground-state-cooled ions yields a Stark shift extinction ratio $\gamma \equiv \delta_{\text{an}}/\delta_{\text{n}} = 10.83(3)$. 1$\sigma$ error bars are smaller than data points.}
    \label{fig:SWprofile}
\end{figure}

After laser cooling and optically pumping the ion into the $\Sp$ state, we measure $\delta$ via spectroscopy on the $\Sp \leftrightarrow |3D_{5/2}, m_j = +1/2\rangle$ with a $\lambda=729$~nm probe pulse detuned $\Delta_{729}$ with respect to the unshifted resonance, applied simultaneously with the SW light. Fig.~\ref{fig:SWprofile}a shows the resulting populations $P_{S_{1/2}}$ remaining in the electronic ground state, measured as a function of $\Delta_{729}$ and the ion position along the axial direction $\x$, for an ion with axial motion initially cooled to the quantum ground state. Fits of the excitation spectrum at each position allow determination of $\delta$, which are shown (with fits to a Gaussian-envelope SW profile) in Fig.~\ref{fig:SWprofile}b. Fit $\delta$ values for the data in panel (a) are shown in black, and additionally for ions initially cooled only to the Doppler limit in purple. 

We define the AC Stark shift extinction ratio in the SW, $\gamma \equiv \delta_{\text{an}}/\delta_{\text{n}}$, as the ratio of the AC Stark shift at the antinode, $\delta_\mathrm{an}$ and at the node, $\delta_\mathrm{n}$. Due to optical coupling to sideband transitions, or equivalently as due to the spatial dependence of the AC Stark shift on the ground state \cite{vasquez2024state}, even for an optical profile with perfect intensity extinction at the SW nodes, $\delta_{\text{n}}$ scales with the wavefunction extent and phonon number of motional modes with mode vector projection along the SW direction (see Appendix~\ref{App:motstate}). As a result, we measure a larger $\gamma$ after ground-state cooling of the axial mode (black points and fits) as compared to simply after Doppler cooling (purple). Near the center of SW profile, the fits indicate $\gamma = 10.83(3)$ for ground-state-cooled ions. Measured phonon numbers and associated wavefunction extent after the cooling employed would result in $\gamma_0 \approx 20$ (see Appendix~\ref{App:motstate}). The difference between $\gamma_0$ and our measured $\gamma$ would be explained by residual carrier coupling at the node with Rabi frequency 9.5$\%$ of the SW antinode. Approximately $20$~nm-level fluctuations in trap position on timescales shorter than those over which we average for the Stark shift measurements would account for the discrepancy between $\gamma_0$ and $\gamma$, which may be due to $E$-field noise and resulting jitter in the equilibrium trap position (see Appendix~\ref{App:motstate}). 

\section{SW Doppler cooling}

We employ the demonstrated SW extinction, as well as stable ion positioning in the UV SW, to explore SW Doppler cooling to phonon numbers below the typical Doppler limit. In our experiment, the free-space RW and the integrated SW beams used for Doppler cooling have the same wave vector direction in the $\x-\y$ plane, allowing simultaneous cooling of all three motional modes (see Appendix~\ref{App:motmodes}). We iteratively scan the RW and SW cooling beams' detuning, intensity, and pulse duration to first minimize the decoherence of Rabi oscillations on a $\lambda=729$~nm transition due to the ion's motion and then the phonon numbers of the $R_2$ mode via a more precise method based on imbalance in motional sideband excitation of a $\lambda=729$~nm optical quadrupole transition \cite{Heatingoftrappedion}. Both the optimized RW and SW detunings are consistent with $\Delta = -\Gamma/2$ given the $P_{1/2}$ natural linewidth $\Gamma \approx \tpi 21.5$ MHz. Finally, with the optimized cooling beam parameters, we measure the resulting phonon number via the sideband imbalance method for all three motional modes.

Prior to cooling within experimental sequences, ions are expected to have axial mode occupancies $\bar n_\mathrm{ax}$ of order 100 due to scattering during the preceeding readout (see Appendix~\ref{App:expsetup}). As reported in Table~\ref{tab:Dopplercooling}, after 1500 \textmu s of RW Doppler cooling the phonon numbers in all three modes approach the standard RW Doppler limit $\simeq\frac{\Gamma}{2 \omega_m}(\frac{1}{2}+\frac{\alpha}{\cos^2 \theta_m})$, for motional frequency $\omega_m$, angle between the cooling beam and motional direction $\theta_m$, and coupling of isotropic spontaneous emission into any motional axis $\alpha=1/3$ \cite{Eschner_Morigi_Schmidt-Kaler_Blatt_2003, Stenholm_1986}.

\begin{figure}[t!]
    \centering
\includegraphics[width=0.45\textwidth]{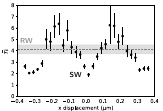}
    \caption{Final $R_1$ mode phonon number after SW Doppler cooling (black points) as a function of axial position. Dashed line and uncertainty indicates the $\bar n$ measured after RW cooling.  Zero displacement corresponds to the central SW node. Error bars indicate 1$\sigma$ uncertainty.}
    \label{fig:SWDC}
\end{figure}

For SW Doppler cooling, we can apply varying strengths of the carrier and sideband interactions by translating the ion through the SW profile. Fig.~\ref{fig:SWDC} shows the measured $\bar n$ in the $R_1$ radial mode following SW after RW Doppler cooling, as a function of trap displacement along the axial direction. In contrast to the SW nodes at which the ion is cooled to below the Doppler limit, at the SW antinodes, the field gradient is zero, which leads to no sideband excitation and only heating from carrier excitation \cite{cirac1992laser}. The regular dependence of $\bar n$ with axial position and the correspondence to the SW periodicity confirm the SW's role in cooling, and the ability to leverage phase-stable delivery for motional control.

For an ion positioned at a SW node, SW Doppler cooling of 700 \textmu s duration with or without preceding RW cooling brings the phonon number in the axial mode to $\sim 1.8\times$ below the RW Doppler limit, with similar enhancement factors in cooling limit observed for the radial modes (Table~\ref{tab:Dopplercooling}). These cooling limits are lower by close to the factor of two predicted within the Lamb-Dicke (LD) regime \cite{cirac1992laser}. The agreement between values of final phonon number $\bar n_\mathrm{ss}$ measured with and without RW Doppler cooling also indicates the effectiveness of SW Doppler cooling even from well beyond the LD regime \cite{xing2025trapped}. 

\begin{table}[]

\begin{tabular}{ccccc}
                                        & RW limit & RW     & SW     & RW \& SW \\ \hline
Axial (1.18 MHz) $\bar{n}_{\mathrm{ss}}$ & 10.6     & 12(2)  & 6.9(6) & 6.4(5)   \\ \hline
$R_1$ (4.45 MHz) $\bar{n}_{\mathrm{ss}}$ & 4.4      & 4.2(5) & 2.5(2) & 2.4(2)   \\ \hline
$R_2$ (5.27 MHz) $\bar{n}_{\mathrm{ss}}$ & 3.7      & 2.7(4) & 1.7(1) & 1.7(1)   \\ \hline
\end{tabular}

\caption{Theoretical and experimental (with 1$\sigma$ uncertainty) final phonon numbers. From left to right, columns correspond to theoretical limit of RW Doppler cooling; and measured $\bar n_\mathrm{ss}$ values after 1500 \textmu s Doppler cooling with the RW, 700 \textmu s with the SW and ion positioned at a node, or both, on all three motional modes.}
\label{tab:Dopplercooling}
\end{table}

\section{Ground-state SW EIT cooling}

Having demonstrated simple cooling dynamics in the phase-stable SW, we turn to ground-state EIT cooling for which enhancements of greater practical significance are predicted in SW configurations \cite{xing2025trapped}.

In our experiment, the three-level system consists of the two $4S_{1/2}$ ground states, and $\ket{4P_{1/2}, m_j = +1/2}$. We employ a free-space $\sigmap$-polarized pump beam with $\Delta_p \approx 2\pi\times 120$ MHz, and approximately $\bm{\hat \pi}$-polarized cooling beams originating either from the integrated SW couplers, or the RW coupler (Fig.~\ref{fig:Ion trap device}) for a direct comparison between RW and SW cooling. Note that the cooling axis in RW EIT is the wave vector difference between the pump and cooling beam; for SW EIT it is simply the SW wave vector itself. Both RW and SW cooling axes have considerable projection along radial and axial modes (see details on mode orientations in Appendix~\ref{App:motmodes}). Here, we tune $\Omega_p$ to optimally cool the $R_1$ mode for both RW and SW EIT cooling, and measure the resulting phonon number in all modes to directly compare the performance of RW and SW EIT in terms of cooling rate, limit, and bandwidth. 

We use two sets of $\Omega_p$ and $\Delta_c$ according to theoretical expectations for RW and SW EIT cooling targeting the $R_1$ mode. We set the AC Stark shift from $\Omega_p$ on $\Sm$ to the expected value using the inverse AC Stark shift sequence described in Appendix D. With $\Omega_c$ below saturation, we also vary $\Omega_p$ and $\Delta_c$ to confirm that the chosen values minimize $\bar n_{\text{ss}}$ for both RW and SW EIT. Lastly, we minimize the cooling duration by increasing $\Omega_c$.

\begin{figure}[t]
    \centering
\includegraphics[width=0.48\textwidth]{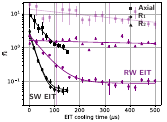}
    \caption{Trajectories of RW and SW EIT cooling. Results of all motional modes use cooling parameters optimized for $R_1$. Corresponding final phonon numbers and cooling rates are listed in Table~~\ref{tab:EITcooling} (see raw data of the sideband imbalance measurement after EIT cooling in the appendix). Error bars indicate 1$\sigma$ uncertainty.} 
    \label{fig:EITcooling}
\end{figure}

Fig.~\ref{fig:EITcooling} shows the measured phonon numbers for all motional modes as a function of cooling duration, for optimal RW and SW EIT cooling parameters chosen to cool the first radial mode $R_1$ with frequency $\omega_{R_1}=\tpi4.45$ MHz, and with the ion positioned at the node of the SW cooling beam in the latter case. We fit the cooling trajectories to $\dot \nbar(t) = -W_c \nbar(t) + R_h$, with $W_c$ the cooling rate and $R_h$ the total heating rate and from which we infer $\bar n_\mathrm{ss} = R_h/W_c$; results of the fits and final $\bar n$ values achieved are summarized in Table~\ref{tab:EITcooling}. RW EIT of 500 \textmu s duration on the target $R_1$ mode has $W_c$ and $\bar n_\mathrm{ss}$ comparable to previously demonstrated results \cite{roos2000experimental, lechner2016electromagnetically}, but negligibly affects the other two motional modes. On the other hand, $\sim$150 \textmu s of SW EIT cooling brings this mode to $\nbar_{\text{ss}} = 0.050(3)$ within a shorter duration. At the same time, SW EIT significantly reduces the phonon numbers in the other two motional modes. In particular, despite both optimized on a mode of $\sim4\times$ higher frequency, RW EIT weakly cools the axial mode while SW EIT brings it to ${\sim}10\times$ below its Doppler limit, clearly demonstrating the SW scheme's higher cooling bandwidth \cite{xing2025trapped}. Due to increased positional drifts observed at high emitted intensities (see Appendix~\ref{App:drift}), we operate at $\Omega_c$ well below that at which the SW EIT cooling rate would saturate, indicating potential for further future improvements in $W_c$.

\begin{table}[]
\begingroup
\setlength{\tabcolsep}{6pt} 
\begin{tabular}{ccccc} 
      & \multicolumn{2}{c}{RW EIT}              & \multicolumn{2}{c}{SW EIT}              \\ \cline{2-5} 
      & $\nbar_{\text{ss}}$ & $W_c$ (ms$^{-1}$) & $\nbar_{\text{ss}}$ & $W_c$ (ms$^{-1}$) \\ \hline
Axial & 5(1)$^*$            &                   & 0.8(1)$^*$          & 25(3)             \\ \hline
$R_1$ & 0.088(5)            & 21(2)             & 0.050(3)            & 57(3)             \\ \hline
$R_2$ & 1.3(1)$^*$          &                   & 0.056(4)            & 52(3)             \\ \hline
\end{tabular}
\caption{Final phonon numbers and cooling rate (with 1$\sigma$ uncertainty) of cooled motional modes after 500 \textmu s RW EIT or 150 \textmu s SW EIT cooling optimized for the $R_1$ mode. For well-cooled modes, we fit cooling trajectories to extract $W_c$ and $\nbar_{\text{ss}}$; for weakly cooled modes, fit results were relatively unreliable, and we report the final measured $\bar n$, marked with asterisks. RW EIT cooling rates of the negligibly cooled axial and $R_2$ modes are omitted due to high uncertainty.}
\label{tab:EITcooling}
\endgroup
\end{table}

Our experimentally observed $\bar n_\mathrm{ss}$ values are significantly above the ${<}10^{-4}$ levels theoretically possible with SW EIT cooling \cite{xing2025trapped}. As detailed in Appendix~\ref{App:EITmodeling}, key experimental limitations to $\bar n_\mathrm{ss}$ in our experiments arise from, in order of importance: (1) $\sigmap$ beam polarization impurity arising from quantization axis oscillations due to trap RF-induced magnetic fields \cite{malinowski2021unitary, joshi2024characterization}, resulting in ${\sim}0.6\%$ relative intensity in the $\hat{\bm{\pi}}$ component in our current system (2) positional fluctuations and/or imperfect SW optical extinction that result in residual intensity at the SW node; and (3) motional heating rates. As quantified in our master equation simulations (Appendix H), these known nonidealities limit the achievable $\bar n_\mathrm{ss}$ to the $10^{-2}$ level achieved experimentally. While our model explains the order of magnitude of our $\bar n_{ss}$ results for both RW and SW experiments, our experimentally measured final phonon numbers are still approximately 2$\times$ higher than those predicted from our simulations in the presence of the known nonidealities, likely due to excess technical noise during EIT experimental sequences not accounted for in our independent estimations (see Appendix H). While future experiments will be required to validate our master equation model at significantly lower phonon numbers, our simulations indicate that polarization impurity in the $\sigmap$ beam are the dominant nonideality, and hence that implementation of RF electrode structures that eliminate oscillations in $B$-field orientation due to trap RF currents would enable SW EIT to reach $\bar n \approx 5\times10^{-3}$ for the radial modes, an order of magnitude below those observed currently. Reduced electric field noise, and therefore heating and positional fluctuations, could allow yet further reductions towards the $\bar n \sim 10^{-5}$ level that is the theoretical limit of the scheme for our parameters, limited by sideband excitations associated with the RW pump beam employed. 

Nevertheless, in the presence of all present experimental nonidealities, our experiments demonstrate simultaneous enhancements in cooling limit, rate, and mode frequency bandwidth due to carrier nulling. The performance realized further compares favorably across metrics to polarization gradient cooling recently realized in integrated platforms \cite{clements2024sub}, indicating the promise of SW EIT schemes for fast, broadband laser cooling in integrated settings. 

\section{Discussion and outlook}

Our work suggests various possible improvements to the operations implemented, along with broader possibilities for atomic control in scalable platforms. With respect to SW EIT cooling specifically, integrated delivery of all required beams, including the $\sigmap$ pump, could be implemented by switching to a $B$-field oriented along the trap surface normal, and a pump beam emitted along the same direction with passive structures emitting pure circular polarization \cite{he2014chip, spektor2023universal, Massai_Schatteburg_Home_Mehta_2022, natarajan2024arrayed}. Trap RF electrode routing that minimizes time-varying $B$-fields perpendicular to the quantization axis would allow SW cooling to reach significantly lower phonon numbers. Operation at higher quantizing magnetic field strengths would further lessen sensitivity to polarization impurities. Straightforward improvements to the fabrication process will further reduce the overall input-to-ion loss, as would higher-efficiency UV gratings based e.g. on higher index-contrast waveguide platforms \cite{jaramillo2025hfo2} and/or multi-layer gratings \cite{knollmann2025collection}. Together with improved transparent conductive shielding, the resulting higher cooling beam intensities would allow yet higher cooling rates than those demonstrated here and thereby lower $\bar n_\mathrm{ss}$ values for a given heating rate, as detailed in Appendices~\ref{App:EITmodeling} and \ref{App:drift}. 

Two-ion crystals cooled to phonon numbers demonstrated in this work would support laser-based entangling gates with errors due to motional occupancy at the $<10^{-4}$ level \cite{ballance2017high}. Achieving such occupancies in a single GS cooling step may alleviate overheads associated with cooling. Fast GS cooling also affects architectural considerations in trapped-ion QC systems, for example in evaluating tradeoffs between laser-based and microwave gates, the latter which can operate at high fidelities with relaxed cooling requirements \cite{hughes2025trapped}. We note that even in the case of microwave gates, efficient GS cooling methods may be essential for further fidelity improvements, robustness against compounding imperfections in large-scale systems, and allowing for ion transport at higher speeds and resulting excitations \cite{bowler2012coherent}.  The sub-recoil limit phonon numbers at the ${<}10^{-4}$ levels fundamentally enabled by the SW EIT scheme demonstrated here can eliminate the need for sideband cooling even for the most stringent motional mode occupancy requirements. 

The methods demonstrated here can be extended both to multi-ion crystals and more complex internal state structures. Ion chains containing two coolant ions \cite{home2009complete, pino2021demonstration} may be cooled in SWs oriented as in this work, with positioning requirements comparable to those of many laser-based geometric phase gates \cite{leibfried2003experimental}. Radial modes of long ion strings with irregular spacings could be straightforwardly addressed either with a SW or Hermite-Gaussian field with nodal line along the axis. EIT cooling has been demonstrated in ions with nonzero nuclear spin and significantly more complex internal structure \cite{feng2020efficient, huang2024electromagnetically}, for which we expect similar advantages from carrier nulling. Evaluation of carrier-nulled EIT cooling's performance for highly excited ions would also be an interesting avenue for future work \cite{bartolotta2024laser}. 

Concepts similar to those demonstrated here may be applied to systems based on individual neutral atoms requiring ground-state cooling \cite{thompson2013coherence,  kaufman2012cooling, jenkins2022ytterbium}, including with more general structured light fields \cite{schmiegelow2012light, verde2023trapped}. The optical addressing and ion positioning precision demonstrated in this work also indicate the practicality of related concepts for quantum logic \cite{mai2025scalable, cui2025transverse, kolhatkar2025efficient, momenzadeh2026individual} in integrated devices, and metrology \cite{Peshkov_HG2023}, suggesting broader potential for scalable optical addressing to enhance atomic control at a basic physical level.

\section*{Acknowledgments}
We thank Floris Falke, Arne Leinse, and the team at LioniX International for fabrication of the trap devices utilized here, and Sonia Garcia Blanco's group at University of Twente for alumina deposition. We thank Orion Smedley, Alex Shi, and Nelson Ooi for contributions to the apparatus, Oscar Jaramillo for assistance with SEM imaging, and Jonathan Home for support in the early design stages of this work. We thank John Chiaverini and Jonathan Home for comments on the manuscript. 

We acknowledge support from an NSF CAREER Award (No. 2338897); IARPA and the Army Research Office, under the Entangled Logical Qubits program, Cooperative Agreement Number W911NF-23-2-0216; the NSF NQVL program (No. 2435382); the Alfred P. Sloan Foundation; and Cornell University. 

The views and conclusions contained in this document are those of the authors and should not be interpreted as representing the official policies, either expressed or implied, of IARPA, the Army Research Office, or the U.S. Government. The U.S. Government is authorized to reproduce and distribute reprints for Government purposes notwithstanding any copyright notation herein.

\section*{Data, code, and materials availability}
Source data and simulation results are available online \cite{data}.

\appendix

\section{Device design and fabrication}
Trap devices were fabricated at LioniX International, with support from the University of Twente for \AlO{} deposition, on 100 mm-diameter wafers. The layer stack used is similar to previous work \cite{mehta2019towards} using this foundry platform, but additionally incorporating \AlO{} waveguides for low-loss light delivery at blue and UV wavelengths, indium tin oxide (ITO) as a transparent conductor to shield windows in the top metal introduced to allow light propagation out of the device \cite{niffenegger2020integrated}, and utilizing gold for both top metal and ground plane layer. All waveguide features were fabricated using electron beam lithography in this work. 

Emitting gratings were designed according to the methods presented in \cite{beck2024grating}, and simulated with full 3D FDTD (Ansys Lumerical FDTD and Tidy3D). The $1\times 2$  397 nm MMI splitter used for the SW generation has a simulated excess loss of 0.21 dB. Simulated grating losses for the designs presented in Fig.~\ref{fig:levstruc_wavelengths} range from 3 dB for the \SiN{} designs for 729 nm wavelength, to 8 dB for present blue/UV wavelengths in \AlO{}. The latter efficiencies are limited by the relatively low refractive index contrast between \AlO{} and \SiO{}; incorporation of higher refractive index contrast core materials for blue/UV waveguides \cite{jaramillo2025hfo2} and/or multi-layer gratings \cite{knollmann2025collection} would enable higher efficiency gratings of the same footprint. 

\begin{figure*}[t!]
    \centering
\includegraphics[width=0.75\textwidth]{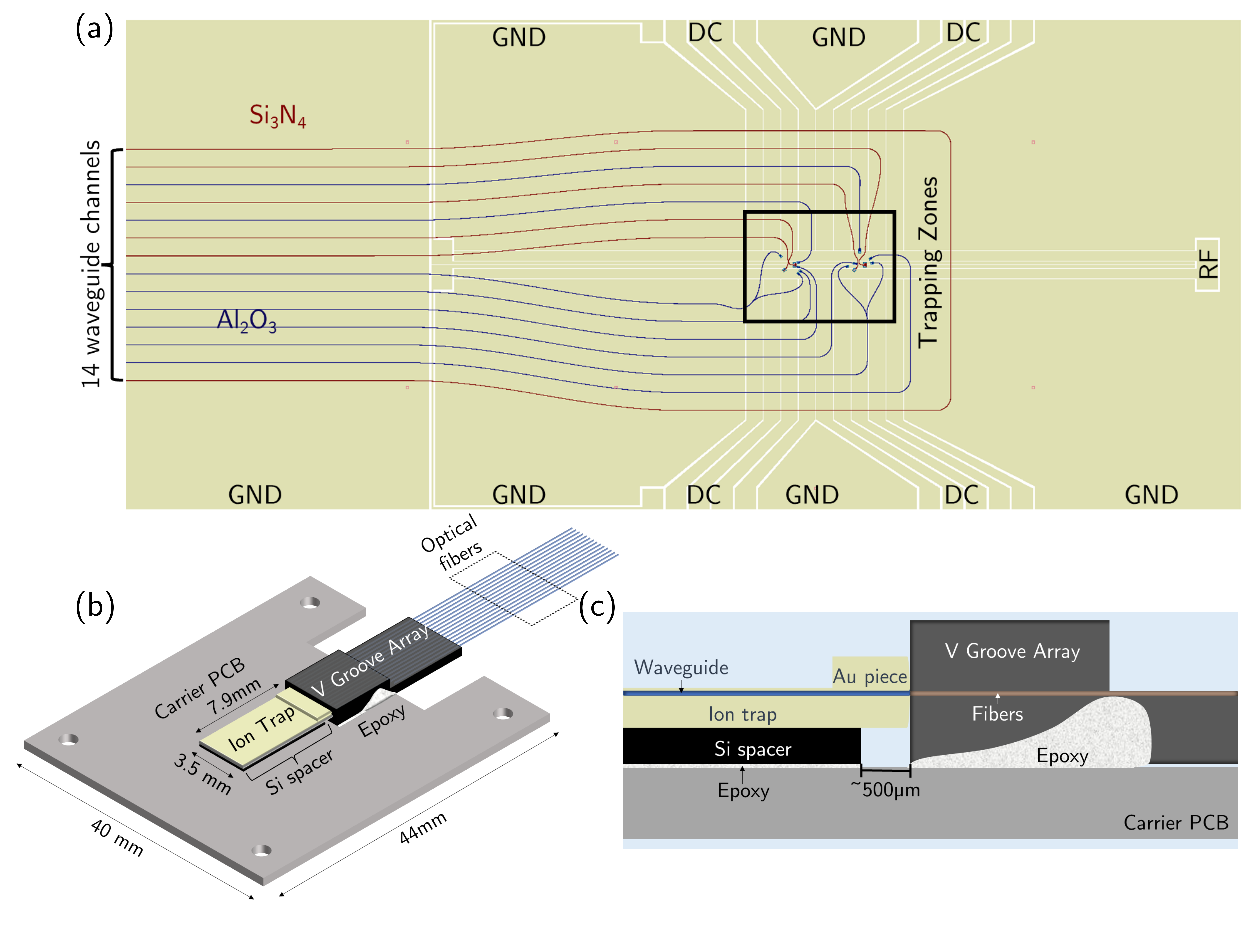}
    \caption{Ion trap chip design and schematic assembly. (a) Design layout of the ion trap device showing \AlO{} and \SiN{} waveguide features, and DC and RF electrode configurations. (b) Schematic of fiber V-groove array (VGA) and trap device mounting on the carrier board fixed to the 4K experimental chamber. (c) Side view of the fiber VGA attachment.}
    \label{fig:fiber attachment}
\end{figure*}

Input couplers to the on-chip waveguides rely on lateral tapers for the \AlO{} waveguides, which taper down to 150 nm width at the fiber-coupling edges (the left edge in Fig.~\ref{fig:fiber attachment}a), to result in weakly confined modes relatively well matched to optical fiber modes. The same vertical taper designs used in \cite{mehta2020integrated} were used for the \SiN{} waveguides. After fabrication of the \AlO{} waveguides on the present devices but prior to the SiO$_2$ cladding deposition, delamination of the narrow waveguides within approximately 300 \textmu m of the intended input coupling chip edge was observed, due to poor adhesion of narrow \AlO{} features to the substrate; this issue was present only where the \AlO{} waveguide features were narrow. As a result, chip die were diced at that facet approximately 350 \textmu m inwards from where originally intended, at such a distance in from the originally intended facet that the lateral taper had expanded to a width that remained fixed to the thermal \SiO{} beneath the waveguides. While this allowed input coupling to all channels, it resulted in somewhat higher edge-coupling losses on all channels than expected, as discussed below. This issue can be straightforwardly avoided by incorporating large-area \AlO{} features off the actual trap die to anchor the narrow laterally tapered waveguides in future fabrication runs.     

\section{Trap assembly, packaging and power handling \label{App:trap}}
On-chip waveguide channels (see Fig.~\ref{fig:fiber attachment}a) are fed by a fourteen-channel V-groove array (VGA), supplied by OZ Optics. The VGA is populated with Fibercore SM300 and Nufern S630-HP fibers with 127 \textmu m pitch, to deliver light to \AlO{} and \SiN{} waveguides, respectively. We use the outer two channels to align the VGA with the trap device by maximizing transmission through the outer \SiN{} waveguide loop visible in Fig.~\ref{fig:fiber attachment}a.

We polish with a method similar to \cite{mehta2020integrated} to reduce roughness from dicing and loss associated with the coupling interface of the ion trap chip. To facilitate polishing, a piece of $1.2\times  3.3$ mm$^2$ SiO$_2$ with 500 \textmu m thickness is coated with 500 nm of gold (to prevent possible stray field from the exposed dielectric) and attached with a conductive epoxy (EPO-TEK, model H21D) to the top surface of the trap chip at the fiber-coupling edge shown in Fig.~\ref{fig:fiber attachment}. Subsequently, we spin-coat the chip in a protective photoresist (FSC M) and mount it in a custom-made holder for facet polishing. Because our devices contain nearly twice as many optical channels as those in \cite{mehta2020integrated}, we employ a polishing holder designed to minimize lateral curvature of the facet and hence the air-gap between fibers and waveguides.

The trap chip's Si substrate thickness (500 \textmu m) differs from that of the VGAs used (1 mm). To match the waveguide and fiber heights, we attach a Si piece of $3.5\times7$~mm$^2$ dimension and 500~\textmu m thickness to the carrier PCB using EPO-TEK T7110 epoxy, as a ``pedestal" on which to mount the trap (see Fig.~\ref{fig:fiber attachment}b). In addition to matching the waveguide height to that of the VGA substrate thickness (1 mm), the epoxy thickness (roughly $ 50$~\textmu m) results in a small gap between the VGA substrate and the PCB, providing clearance to position the VGA for fiber attachment. Subsequently, we scribe the Si substrate of the die to break through the native oxide and remove the protective photoresist. We attach the die to the Si pedestal using EPO-TEK 301-2FL epoxy (epoxy without filler) to minimize the gap between the Si pedestal and the trap chip. Furthermore, we leave an approximately 500 \textmu m distance (see Fig.~\ref{fig:fiber attachment}c) between the fiber coupling edge of the trap die and the edge of the Si spacer on the fiber coupling side, which is crucial to prevent epoxy from wicking between the VGA and the trap interface during VGA attachment. We apply a drop of conductive silver epoxy (EPO-TEK, model H21D) on both sides of the trap chip near where the native oxide was scribed to contact the Si substrate, gold-coated SiO$_2$ piece, and the ground of the trap metal to the ground of the carrier PCB. We wirebond the trap electrodes to the carrier PCB (gold electrodes) using 25~\textmu m diameter gold wire with a wedge wirebonder.

To attach the fiber array, we clamp the carrier PCB in a custom mount heated to 90$^\circ$C, 10$^\circ$C above the glue curing temperature. The VGA is clamped to a Thorlabs HFA001 fiber array holder and mounted on a 6-Axis NanoMax Stage (Thorlabs MAX603D) for automatic alignment by maximizing the transmission through the loop waveguides shown in Fig.~\ref{fig:fiber attachment}a. We use three degrees of freedom (DOF) for automatic alignment, and the other three DOFs are left for manual alignment. After alignment, we press the VGA against the chip so that static friction between the VGA and trap interfaces allows us to passively maintain coupling for multiple hours. 

From the total loop transmission, we measure an insertion loss of approximately 5.5 dB/facet at $\lambda=729$ nm, in comparison to the $\sim 1.5$ dB level loss measured for the same taper couplers in \cite{mehta2020integrated}. We attribute this excess loss to the 350 \textmu m indicing of the trap devices mentioned above, which resulted in an optical mode more highly confined in the \SiN{} at the facet and less matched to that of the fiber; the same effect resulted in somewhat higher insertion losses than designed for the \AlO{} channels. We manually apply drops of EPO-TEK T7110 epoxy to both sides of the VGA (see Fig.~\ref{fig:fiber attachment}c), which wicks into the gap between the VGA and the carrier PCB, with the gap between the VGA and the bottom Si spacer preventing the epoxy from wicking into the interface between the trap die and the VGA. We release the VGA after two hours of curing, while the carrier PCB is kept overnight on the heated mount to minimize the risk of uncured epoxy. We measure additional losses of 0.5 dB and 0.14 dB per facet (at 729 nm) during cool-down from 90$^\circ$C to room temperature and from room temperature to 4.8 K, respectively. We attached and cooled down three devices in the same manner, and measured similar insertion loss and cool-down behavior. Additionally, we do not observe increased losses after further thermal cycles between room temperature and 4K.

Prior to the eventually employed method, we fiber-attached several samples following the procedure described in \cite{mehta2020integrated}, where epoxy was applied at the VGA–trap interface. Even though the polished facets (smooth at the single nanometer level) are in mechanical contact, at 375 nm, we observed photo-degradation at the fiber feedthrough for input powers as low as 25~\textmu W. This effect occurred consistently across all tested devices, regardless of the epoxy type (transparent or opaque) or the polishing method. In contrast, the method described above involving no epoxy in the optical path substantially improved UV tolerance, allowing up to 15 mW of 375 nm light (the maximum available laser power) to be delivered without detectable degradation.

\section{Waveguide/grating optical characterization and UV waveguide loss \label{App:waveguideloss}}

We profile beams emitted by the various gratings (Fig. 2c) using a microscope mounted on a vertical motorized translation stage (Newport TRB12CC). The stage provides 0.25 \textmu m unidirectional positioning accuracy and allows us to image grating emission at different heights above the trap device. The emission was collected through a 0.95 NA objective (Olympus MPLANAPO50x) and imaged onto a scientific CCD camera (Lumenera Infinity 3S-1UR) \cite{mehta2017precise}. We optimize the input polarization by monitoring the grating emission angle roughly 50 \textmu m above the device. TE polarization emits at a higher angle relative to the normal, owing to the larger effective index in \AlO{} and \SiN{} waveguides. We measure a targeting inaccuracy below 2 \textmu m for wavelengths spanning 375 to 866 nm. In addition, we measure waveguide loss of 1.7(3) dB/cm at $\lambda=397$ nm for quasi-TE modes of single-mode \AlO{} waveguides. \SiN{} propagation losses were similar to those in \cite{mehta2020integrated}. 

We infer the total optical loss in the 397 nm SW  channel from fiber feedthrough to the ion location by measurement of the differential AC Stark shift of the $\Sp \leftrightarrow \ket{3D_{5/2},m_j={+1/2}}$ quadrupole transition, similar to the measurements presented in Fig.~\ref{fig:SWprofile}. The $\lambda = 397~$nm light is 10 GHz red-detuned from the $S_{1/2}\leftrightarrow P_{1/2}$ transition, such that the quadrupole transition shift is dominated by the AC Stark shift of the $\Sp$ level. At the SW anti-node, we measure $\delta = \tpi 8.04 $~kHz for 161 $\mu$W power input at the vacuum fiber feedthrough. We calculate the laser intensity $I_\text{SW}$ at the anti-node corresponding to this AC stark shift from the $S_{1/2}\leftrightarrow P_{1/2}$ dipole transition strengths, and the power emitted in the SW according to the expression $P_{\mathrm{tot}}=\frac{1}{4}I_\text{SW} \pi w_{\mathrm{t}} w_{\mathrm{l}}$. The transverse and longitudinal beam waists $w_{\mathrm{t}} = 8.95$ \textmu m and $w_{\mathrm{l}} = 5.75$ \textmu m, respectively, are obtained from the beam profile measurements discussed above. Through these observations we infer a total loss of 33.3 dB. 

This total loss includes fiber connector losses, fiber-to-chip edge-coupling loss, waveguide loss, MMI loss, grating inefficiency, and fiber-mode concentricity; estimates of these contributions are shown in Table~\ref{tab:loss_budget}. We find the measured total loss to be $\sim10$ dB higher than accounted for by these contributions. Potential delamination in the \AlO{} waveguide despite the in-dicing may possibly account for the additional loss (this is not directly visible in the fabricated samples coated with metals). 

\begin{table}[t!]
\begingroup
\setlength{\tabcolsep}{6pt} 
\centering
\begin{tabular}{| l  | l |}
\hline
\textbf{Loss source} & \textbf{Loss (dB)} \\ \hline
Edge coupling (simulated) & 3 dB \\ \hline
Additional in-dicing loss (simulated) & 4 dB \\ \hline
Fiber connectors & $\sim$4 dB\\ \hline
Grating efficiency (simulated) & 8 dB\\ \hline
Waveguide loss (measured) & 0.7 dB \\ \hline
MMI splitter (simulated) & 0.21 dB\\ \hline
Beam misalignment (measured) & 0.5 dB\\ \hline
Fiber core concentricity (simulated) & 0 - 2.8 dB \\ \hline
\textbf{Total} & $\mathrm{20.4 - 23.2}$  dB \\ \hline
\end{tabular}
\caption{Approximate loss budget for the $\lambda=397$~nm channel feeding the SW profile. The simulated and measured loss sources account for 20.3 - 23.1 of the observed 33.3 dB total fiber-to-ion loss; we attribute the remaining ${\sim}10$ dB to potential effects of delamination of the narrow input coupling waveguide near edge-coupling facet which occurred on this particular fabrication run.}
\label{tab:loss_budget}
\endgroup
\end{table}

\section{Laser systems, cryogenic apparatus, and trap operation \label{App:expsetup}}
Ion trap devices are operated at cryogenic temperature (4.8 K) to reduce electric field noise, enhance ion lifetime, and enable rapid reconfiguration. We employ a closed-cycle Gifford-McMahon cryocooler, which is ceiling-mounted and mechanically decoupled from the experiment chamber hosting the trap device to minimize vibrations. The trap chip is mounted in a copper chamber with thick $25.4$-cm walls to shield time-varying magnetic fields. A static magnetic field of 8.9~G is generated by permanent magnet rings installed directly the 4 K chamber, generating a quantizing field with simulated fractional uniformity on the level of $10^{-4}$/mm near the device center.

We use solid-state diode lasers and frequency doublers (supplied by Toptica Photonics) to deliver all wavelengths except $\lambda=729$~nm. These diode lasers are frequency-stabilized via a wavelength meter (HighFinesse WS7-30), with the exception of the $\lambda=375$~nm photoionization laser which is free-running. The $\lambda=729$~nm light is generated by a CW Titanium–Sapphire laser (M-Squared SolsTiS), which is locked to a high-finesse reference cavity (Stable Laser Systems). We measure $T_2$ decay times on the optical quadrupole transition of over 20~ms at present, such that decoherence on this transition negligibly affects the results presented in this work.

Neutral Ca atoms are supplied to the 4 K chamber via a resistively heated oven mounted in the room-temperature vacuum chamber. The $\lambda=422$ nm photoionization beam, 729 nm qubit beam, and 854 and 866~nm repump beams are delivered through integrated optics, along with both the 397 nm SW and RW EIT $\hat{\bm{\pi}}$-polarized cooling beams. We employ a free-space optical path for a $\lambda=397$~nm beam used for both RW Doppler cooling and qubit state readout; the $\hat{\bm{\sigma}}_{+}$ beam used for state-preparation and as the EIT pump beam; and the 375 nm photoionization beam. Although $\lambda=375$~nm waveguides are present on the chip, this photoionization beam is delivered in free space due to the power required and insertion losses in the present device. A custom fiber feedthrough, similar to that described in \cite{zhang2021scalable,mehta2020integrated} is used to route fibers into the vacuum chamber and to the VGA attached to the chip. 

\begin{figure}[t!]
    \centering
\includegraphics[width=0.35\textwidth]{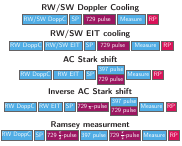}
    \caption{Experiment sequences used for various measurements. SP and RP stand for state prep and repump respectively. }
    \label{fig:exp_sequence}
\end{figure}

Experiment sequences used for various measurements in this work are shown in Fig.~\ref{fig:exp_sequence}, with the wavelengths and pulse times detailed in Table~~\ref{tab:exp_sequence}. 

Each experiment sequence begins after the previous sequence's readout. Phonon number excitation at the beginning of each shot due to scattering during the preceeding readout can be upper bounded by $\Delta\nbar = \eta_0^2\Gamma t_{\text{read}}/2$. For our axial mode's $\eta_0 = 0.164$ and given our measurement duration $t_\text{read}=250$ \textmu s, $\Delta \nbar \approx 450$. Thus, we estimate the axial phonon number to be of order $\sim100$ before Doppler cooling as we use a close-to-saturation readout pulse.

\begin{table}[t!]
\begin{tabular}{ccccc}
          & Pulse time (\textmu s) & 397 nm              & 866 nm  & 854 nm  \\ \hline
RW DoppC     & 1500                & FS                  & $\surd$ &         \\ \hline
SW DoppC     & 700                 & SW                  & $\surd$ &         \\ \hline
RW EIT    & 500                 & FS $\sigmap$ and RW & $\surd$ &         \\ \hline
SW EIT    & 150 to 200          & FS $\sigmap$ and SW & $\surd$ &         \\ \hline
SP        & 2 to 5              & FS $\sigmap$        & $\surd$ & $\surd$ \\ \hline
Measure   & 250                 & FS                  & $\surd$ &         \\ \hline
397 pulse & varying             & SW or FS $\sigmap$  &         &         \\ \hline
Repump        & 5                   &                     & $\surd$ & $\surd$ \\ \hline
\end{tabular}
\caption{Details of the pulses employed in the sequences shown in Fig.~\ref{fig:exp_sequence}. Here, FS, RW, SW stand for free-space, integrated RW coupler, and integrated SW coupler beam delivery, respectively. All instances of the ``729 pulse" in Fig.~\ref{fig:exp_sequence} correspond to $\lambda  = 729$ nm light sent through the integrated coupler. }
\label{tab:exp_sequence}
\end{table}

We utilize voltage sets applied to the DC electrodes of the trap to control axial confinement, displacements along radial directions for micromotion-compensation, radial mode orientation, and axial displacement within the SW. These voltages are calculated following the methods in \cite{allcock2010implementation} based on simulations of the trap geometry in Nullspace ES \cite{nullspace}, and applied via an open-source DAC (Fastino). Radial confinement is provided by an RF signal generator (Rigol DSG800), impedance-matched to the trap with a lumped-element resonator circuit \cite{gandolfi2012compact}. 

Stray DC fields are compensated using sinusoidal modulation of the pseudopotential in a ``tickle" technique \cite{Ibaraki_Tanaka_Urabe_2011}, probing  parametric driving of the secular motion as an indication of displacement of the ion from the RF null. Stray fields are compensated with the voltage sets described above. Since the SW has an intensity profile that varies significantly along the horizontal RF micromotion direction ($\bm{\hat y})$, accurate micromotion minimization via stray-field compensation is critical for ion positioning at the node. 

\begin{figure}[t!]
    \centering
\includegraphics[width=0.48\textwidth]{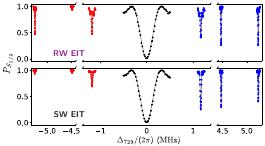}
    \caption{Sideband spectroscopy after RW (top) and SW (bottom) EIT cooling. Measurements are taken with pulse times of 2.8, 30, and 60 \textmu s for the carrier, axial sidebands, and radial sidebands, respectively. Imbalance in RSB and BSB amplitudes are used to infer $\nbar$ values reported in Fig. \ref{fig:EITcooling} \cite{Heatingoftrappedion}. Error bars indicate 1$\sigma$ uncertainty.} 
    \label{fig:EIT_spec}
\end{figure}

For determination of $\Omega_p$ parameters that induces the target $\delta_{\text{EIT}}$ for both RW and EIT cooling, we use the AC Stark shift sequence in Fig.~\ref{fig:exp_sequence} to measure the AC Stark shift of the EIT pump beam on $\Sm$. A Ramsey measurement is used to quantify the small AC Stark shift of the $\Sp$ state induced by impurities in the pump beam. These two results are used to infer the $\sigmap$ pump beam's purity and the associated limit to cooling, detailed in Appendix G. Fig.~\ref{fig:EIT_spec} shows examples of the $\lambda = 729$ nm sideband spectroscopy after RW or SW EIT cooling sequences that are used to extract the phonon number following optimized cooling. 

We measure compensation position drifts of a few nanometers per minute when light is sent through the integrated SW gratings couplers, but observe that these photo-induced stray fields generally relax back to their initial values over several hours (see details on observed drifts in Appendix~\ref{App:drift}). Compensation fields remain approximately stable (within a few percent) over weeks of normal operation. Conductive ITO shielding of the exposed dielectric windows likely contribute to this stability in the presence of the various integrated blue and UV beams; however a more rigorous study is needed to fully understand the effect of ITO in this context.

\section{Motional mode orientations and cooling coefficients \label{App:motmodes}}
\begin{figure}[t!]
    \centering
\includegraphics[width=0.48\textwidth]{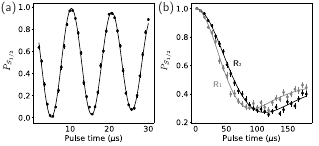}
    \caption{(a) Carrier and (b) BSB flopping with the same input power of $\lambda = 729$ nm light. We performed the RW Doppler cooling sequence shown in Fig.~\ref{fig:exp_sequence} before the flopping. Error bars in both plots indicate 1$\sigma$ standard errors from projection noise.}
    \label{fig:mode_coupling}
\end{figure}

To allow simultaneous cooling and probing of $R_1$ and $R_2$, we apply voltage sets to the trap's DC electrodes that allow for rotations of the radial mode orientations \cite{allcock2010implementation}. We measure the radial modes' orientation from their coupling strength to the $\lambda=729$~nm beam, whose emission angle is $\theta = 34 \degree$ to the trap surface normal. In Fig.~\ref{fig:mode_coupling}, we fit the carrier Rabi oscillations to 
\begin{equation}
P_{\downarrow}(t) = 1-\sum_{n=0}^\infty P(n)\sin^2{\left( \frac{\Omega_0}{2}(1-n(\eta_0\sin\theta)^2 t\right)}
\end{equation} 
to extract $\Omega_0 = 0.628(2)$ \textmu s$^{-1}$. BSB Rabi oscillations are fit to 
\begin{equation}
P_{\downarrow}(t) = 1-\sum_{n=0}^\infty P(n) \sin^2{\left( \frac{\Omega_0}{2}\eta_0\cos \theta\cos\phi\sqrt{n+1}{} t\right)}. 
\end{equation} 
where $P(n) = \frac{\nbar^n}{(\nbar+1)^{n+1}}$ are the thermal-state populations and $\eta_0 = \frac{2\pi}{\lambda} \sqrt{\hbar/(2 m \omega_m)}$ the Lamb-Dicke parameter for wavelength $\lambda$, ion mass $m$, motional mode frequency $\omega_m$. These fits allow us to extract the radial mode rotation angles from the trap normal of $\abs{\phi} = 51(2)\degree$ and $49(1)\degree$ for modes $R_1$ and $R_2$ respectively. 

The result is approximately consistent with both $R_1$ and $R_2$ being rotated $45^\circ$ counterclockwise with respect to the $\z$ and $\y$ axes respectively. As a result, for our analysis we take normal vectors associated with the modes to be $\hat{\bm{n}}_\mathrm{R1} = \frac{1}{\sqrt{2}}(-\hat{\bm{y}}+\hat{\bm{z}})$, $\hat{\bm{n}}_\mathrm{R2} = \frac{1}{\sqrt{2}}(\hat{\bm{y}}+\hat{\bm{z}})$. 

\begin{table}[t!]
\begingroup
\setlength{\tabcolsep}{4.5pt} 
\begin{tabular}{l l l l l l l l l l }
      & RW $\kappa$ & SW $\kappa$ &  &  &          &             &              &              &              \\ \cline{1-3}
Axial & 1.425       & 0.612       &  &  &          & pump        & RW           & SW1       & SW2       \\ \cline{1-3} \cline{6-10} 
$R_1$    & 0.511       & 0.433       &  &  & $\theta$ & 90$\degree$ & 60$\degree$  & 60$\degree$  & 60$\degree$  \\ \cline{1-3} \cline{6-10} 
$R_2$    & 0.196       & 0.433       &  &  & $\phi$   & 45$\degree$ & 124$\degree$ & -45 $\degree$ & 135$\degree$
\end{tabular}
\caption{(left) EIT cooling coefficients $\kappa$, for the RW configuration: $|\hat{\bm n} \cdot (\bm k_p-\bm k_{\text{RW}})/k_0|$ and the SW configuration: $|\hat{\bm n} \cdot (\bm k_{\text{SW1}}-\bm k_{\text{SW2}})/k_0|$. Motional mode unit vectors for each mode are $\hat{\bm n}_\text{ax} = (1,0,0)$, $\hat{\bm n}_\text{R1} = (0,-\frac{1}{\sqrt{2}},\frac{1}{\sqrt{2}})$, $\hat{\bm n}_\text{R2} = (0,\frac{1}{\sqrt{2}},\frac{1}{\sqrt{2}})$, and $\bm k = k_0(\sin\theta \cos \phi, \sin\theta \sin \phi, \cos \theta)$, according to the axes defined in Fig~\ref{fig:Ion trap device}c. (right) Propagation angles for the involved EIT beams.}
\label{tab:coupling}
\endgroup
\end{table}

For both RW and SW EIT cooling, $W_c$ is proportional to $\kappa^2\eta_0^2\Omega_c^2$ for low cooling beam Rabi frequency $\Omega_c$. $\kappa$ denotes a ``cooling coefficient" that accounts for the motional mode's normal vector relative to the axis cooled, defined by the relevant $\bm k$-vectors. These $\kappa$ values for both RW and SW EIT cooling of each motional mode are defined and listed in Table~\ref{tab:coupling}; given our mode and beam orientations, SW EIT cooling acts on all modes, whereas RW EIT would cool the $R_1$ and axial modes, with much reduced coupling to the $R_2$ mode.

\section{AC Stark shift motional-state dependence \label{App:motstate}}

To profile the $\lambda=397$ nm SW using the ion as a high-resolution probe, we used the AC Stark shift sequence illustrated in Fig.~\ref{fig:exp_sequence}. After cooling during each experiment sequence, the $\sigmap$ beam is tuned near to resonance with the $\ket{4S_{1/2}} \leftrightarrow \ket{4P_{1/2}}$ transition to optically pump population into $\Sp$. Then, we measure the spatially dependent AC Stark shift $\delta$ through spectroscopy on the $\Sp \leftrightarrow |3D_{5/2}, m_j = +1/2\rangle$ transition while sending $\lambda= 397$ nm light detuned $\Delta=-1$ GHz from the $S_{1/2} \leftrightarrow P_{1/2}$ transition through the SW channel. In the experiment, we use a fiber polarization controller to adjust the input polarization such that $\lambda= 397$ light couples into the waveguide's quasi-TE mode and emits with approximately $\hat{\bm\pi}$-polarization. 

\begin{figure}[t!]
    \centering
\includegraphics[width=0.48\textwidth]{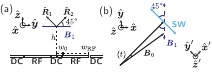}
    \caption{Geometric considerations relevant to mode orientations and oscillating magnetic fields induced by trap RF currents. (a) Side view of the ion on top of the trap surface. The asymmetric voltage sets, applied on the DC electrodes, rotate the $R_1$ and $R_2$ modes $45\degree$ clockwise. (b) RF current charging and discharging the RF electrodes flows along the trap axis $\x$, resulting in an oscillating magnetic field $\bm{B}_1$ along $\y$ at ion's position. To describe quantization axis oscillations, we define a coordinate system with $\x'$ aligned with the static quantizing magnetic field $\bm B_0$ generated by the permanent magnets.}
    \label{fig:exp_config}
\end{figure}

The ion's secular motion within the SW profile affects the measured AC Stark shift, reducing the contrast between the maximum and minimum AC Stark shift. The ideal TE-polarized SW in Fig.~\ref{fig:exp_config} has
\begin{eqnarray*}
    \bm E (\bm{r,t}) = \bm {\hat n}E_0 \sin{(k_x x + k_y y +\varphi)} \exp\left[i (k_z z - \omega_Lt)\right] 
\end{eqnarray*}
with $\bm{\hat n} = (\bm{\hat x} + \bm{\hat y})/\sqrt 2$, $k_x = k_0\sin{\theta}\cos{\phi}$, $k_y = k_0\sin{\theta}\sin{\phi}$, $k_z = k_0\cos{\theta}$, and $\theta=60\degree$ and $\phi = 45 \degree$ (Table~\ref{tab:coupling}). 

Because the SW is predominantly $\hat{\bm\pi}$-polarized and the laser detuning $\Delta$ in our Stark shift measurements satisfies $\Gamma \ll \Delta \ll \omega_{f, P_{1/2}}$ with $\omega_{f, P}$ the $4P$ levels' fine structure splitting, we approximate the AC Stark shift on $\ket g \equiv \Sp$ in terms of its coupling simply to $\ket e \equiv \Pp$. 

In a frame rotating with laser frequency $\omega_L$ and after the usual optical rotating wave approximation, the dipole interaction strength is $\langle e| \tilde V_\mathrm{dip} |g\rangle = \frac{\Omega(\tilde{\bm{r}})}{2} = \frac{q\bm E(\tilde{\bm r})\cdot \bm d_{eg}}{2}$, with dipole matrix element $\bm d_{eg}$ (in this section we label operators with tildes). The resulting AC Stark shift on $\Sp$ for large laser detuning $\abs{\Delta}$ is
\begin{equation}\label{eq:ac_stark}
\delta \equiv \expval{ \frac{\abs{\Omega(\tilde{\bm{r}})}^2}{4\Delta}}
\end{equation}
with $\tilde{\bm{r}}$ the ion's position operator and the expectation value $\expval\cdot$ corresponding to expectation over the motional states of the ion in the spatially varying profile.  

To leading orders in the LD parameters of all motional modes, from Eq.~\ref{eq:ac_stark} we find for the AC Stark shifts at the node and antinode: 
\begin{eqnarray*}
    \delta_{\text{n}} &&= \frac{|\Omega_0|^2}{4\Delta} \left(k_x^2 \langle \tilde x^2 \rangle + k_y^2 \langle \tilde y^2 \rangle \right) = \frac{|\Omega_0|^2}{4\Delta}\mu\\
    \delta_{\text{an}} &&= \frac{|\Omega_0|^2}{4\Delta} \left(1-k_x^2 \langle \tilde x^2 \rangle - k_y^2 \langle \tilde y^2\rangle \right) = \frac{|\Omega_0|^2}{4\Delta}(1-\mu)
\end{eqnarray*}
where
\begin{equation}
    \mu \equiv  \eta_\text{ax}^2 (2\nbar_\text{ax} +1) + \sum_{i=1,2}\eta_{R_i}^2(2\nbar_{R_i} +1).
\end{equation}
Given the motional modes' orientations as discussed previously, $\tilde x = a_{0\text{ax}} \left( \tilde{a}_{\text{ax}}+\tilde{a}^{\dag}_{\text{ax}} \right)$ and $\tilde y = \frac{1}{\sqrt2}(a_{0R_1}(\tilde{a}_{R_1}+\tilde{a}^{\dag}_{R_1})-a_{0R_2}(\tilde{a}_{R_2}+\tilde{a}^{\dag}_{R_2}))$ are the corresponding position operators of the ion. Here $a_{0j} =\sqrt{\frac{\hbar}{2m \omega_j}}$ for mode $j$ of frequency $\omega_j$; $\eta_j$ is the corresponding Lamb Dicke parameter $\kappa \eta_o$, and $\nbar_j$ is the expectation phonon number for $j \in \{\mathrm{ax}, R_1, R_2\}$. Given the wavepacket extent's influence on the AC Stark shifts, to maximize extinction ratio $\gamma = \delta_{\text{an}}/\delta_{\text{n}}$ and thereby most accurately bound the \emph{optical} extinction via this AC Stark shift profiling, we apply a burst of RW EIT cooling optimized to cool the axial mode. This results in $\bar n$ of $0.53(8)$, $6.0(8)$, and $3.6(4)$ for the axial, $R_1$, and $R_2$ modes respectively. This corresponds to $\mu = 0.047$, which would give an extinction ratio limited by motional occupancies of $\gamma_0 \approx 20$.

The difference between $\gamma$ and $\gamma_0$ may be due to positional fluctuations in the equilibrium trap position resulting from electric field noise and drift \cite{Heatingoftrappedion}, on timescales shorter than those over which we average (of order multiple seconds) to measure AC Stark shifts. These fluctuations are expected to be dominant along the axial direction, which has the lowest trap frequency. Such fluctuations result in AC Stark shifts at the SW node and antinode of $\delta_{\text{n}} = \frac{|\Omega_0|^2}{4\Delta} (\mu + \chi)$ and $\delta_{\text{an}} = \frac{|\Omega_0|^2}{4\Delta}(1-\mu-\chi)$, where $\chi \equiv k_x^2\langle \delta_x^2\rangle$, where $\expval{\delta_x^2}$ is the classical mean squared displacement of the equilibrium trap position. From the known $\mu$ due to secular mode occupancy and our measured extinction $\gamma$, we estimate $\chi \approx 0.037$. Considering the emission angle $\theta$ and SW's angle with respect to the axial direction $\phi$, this suggests an axial RMS variation $\sqrt{\langle \delta_x^2\rangle}=\sqrt{\chi}/(k_0 \sin{\theta} \cos{\phi}) = 19.8$ nm. In simulations, we model the effect of the resulting residual field sampled by the ion at the node by including additional carrier coupling with amplitude $\Omega_{\text{res}} = \Omega_0 \sqrt{\chi}/2 $.

We roughly estimate the plausibility of $E$-field noise as the origin of our observed extinction by extrapolating the $E$-field noise magnitude measured at the secular motional frequencies. The axial mode at $\omega_\mathrm{ax} = 1.18$ MHz has a heating rate $\dot{\nbar} = 3.9(3)\times 10^3~\text{s}^{-1}$, and the relative heating rate for the $R_1$ mode (Fig.~\ref{fig:noise_drift}) supports roughly $S(\nu)\propto \nu^{-2}$ scaling \cite{Sedlaceknoise}. We thus suppose that the power spectral density of the electric field noise can be written over an appropriate frequency range as 
\begin{equation}
S(\nu) = \frac{4m h \omega_\mathrm{ax} }{q^2} \frac{\omega_\mathrm{ax}^2}{\nu^2}\dot{\bar{n}}.
\end{equation}
We find an RMS displacement from $E$-field noise $q\sqrt{\int_{\omega_\mathrm{lo}}^{\infty} S(\nu)d\nu}/(m \omega_\mathrm{ax}^2)$ of approximately 20 nm, consistent with the AC Stark shift observations, if we extrapolate over frequency to $\omega_\mathrm{lo} \sim 1~\mathrm{kHz}$ (each AC Stark shift is measured averaged over timescales of multiple seconds). Though we do not know the low-frequency scaling of $S(\nu)$, this indicates the potential plausibility of $E$-field fluctuation as the origin of the limited extinction, with an effective cutoff of order ${\sim}1\ \mathrm{kHz}$ roughly consistent with some previous estimates \cite{talukdar2016implications}. Measurements of $E$-field noise near ion traps at frequencies well below the secular frequencies would be valuable to better understand this behavior, particularly important in contexts requiring precise positioning of ions with respect to control fields. 

We note that the above is an upper bound on possible positional fluctuations, as it neglects other potential contributors to imperfect extinction. Quasi-TM polarization components of the waveguide modes feeding the SW couplers, and uneven power delivery from the two SW coupler could also worsen the extinction ratio, but are expected to be negligible in our experiment. After optimizing for quasi-TE polarization, we consistently measure the same extinction ratio, suggesting an insignificant TM component. Through optical measurements utilizing a camera focused on different $\z$ positions, we also measure the beam power emitted from the two SW couplers separately on the same device used in the ion trap experiment. The power emitted by the two SW couplers was found to be less than $4\%$ different, which would result in $<0.1\%$ fractional intensity at the SW nodes.

\section{$\sigmap$ beam purity and trap RF-induced $B$-field oscillations}
A significant limitation to final phonon number is due to polarization impurity of the $\sigmap$ beam due to magnetic field oscillations associated with the RF currents flowing in the trap \cite{joshi2024characterization}. As illustrated in Fig.~\ref{fig:exp_config}b, the resulting oscillating magnetic field $\bm B_1$ points along a direction different than that of our static quantizing field $\bm B_0$, therefore resulting in a time-varying quantization axis. 

\begin{figure}[t!]
    \centering
\includegraphics[width=0.48\textwidth]{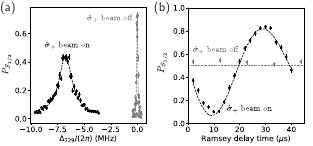}
    \caption{$\sigmap$-beam intensity and polarization purity diagnostics. (a) Using the inverse AC Stark shift sequence with and without the $\sigmap$-polarized pump beam applied during the 729 nm pulse, spectroscopy on $|4S_{1/2}, m_j = -1/2\rangle \leftrightarrow |3D_{5/2}, m_j = +1/2\rangle$ measures the AC stark shift due to the pump $\delta_p$; such measurements are used to set the pump beam strength for RW and SW EIT, with double the $\delta$ for our SW EIT configuration. (b) At the same intensity that resulted in $\delta_p = 2\pi \times 6.75(2)$ MHz in panel (a), a Ramsey sequence on $|4S_{1/2}, m_j = +1/2\rangle \leftrightarrow |3D_{5/2}, m_j = +1/2\rangle$ with and without $\sigmap$ during the Ramsey delay time, indicates $\delta_m = 25.4(2)$ kHz. The second Ramsey pulse is 90 degree phase-shifted, resulting in $P_{S_{1/2}} = 0.5 $ with no frequency shift during the delay time. Error bars in both plots indicate 1$\sigma$ standard errors from projection noise.}
    \label{fig:sigmap_imp}
\end{figure}

To quantify the resulting impurity, we measure the AC Stark shifts $\delta_m$ and $\delta_p$ on $\Sm$ and $\Sp$, respectively, at the same $\sigmap$ beam power. Given the $B$-field oscillations responsible for polarization impurity in our setup as described below, $\delta_p$ arises primarily from $\hat{\bm \pi}$ polarization components. Using the same pump beam frequency as in EIT sequence, $\Delta_m= 120$ MHz blue detuned from the $\Sm \leftrightarrow \Pp$ and $\Delta_p = 95$ MHz from $\Sp \leftrightarrow \Pp$ at $B_0 = 8.9$ G. In Fig.~\ref{fig:sigmap_imp}a, the $\delta_m$ on the level of MHz is measured using the inverse AC Stark shift method in Fig.~\ref{fig:exp_sequence}, where a 729 nm $\pi$-pulse transports the population from $\Sp$ to $|3D_{5/2}, m_j = +1/2\rangle$ before measuring the AC Stark shift $\delta$ on the $|4S_{1/2}, m_j = -1/2\rangle \leftrightarrow |3D_{5/2}, m_j = +1/2\rangle$ transition. On the other hand, we use the Ramsey measurement in Fig.~\ref{fig:exp_sequence} to measure and minimize $\delta_p$ to the level of tens of kHz \cite{lechner2016electromagnetically}. By optimizing the orientations of $\lambda/2$ and $\lambda/4$ waveplates in the $\sigmap$ beam path, we achieve $\delta_p/\delta_m=0.376(3)\%$ in Fig.~\ref{fig:sigmap_imp}b. Accounting for the relative dipole coupling matrix elements, we infer a $\sigmap$ beam impurity of
\begin{eqnarray*}
    \epsilon_{\text{imp}}&&\equiv P_{\hat{\bm{\pi}}}/P_{\sigmap}\\
    &&= 2\Delta_p/\Delta_m \cdot  \delta_p/\delta_m= 0.595\%,
\end{eqnarray*}
where $P_{\hat{\bm\pi}}$ and $P_{\sigmap}$ are the beam power in $\hat{\bm\pi}$ and $\sigmap$ respectively.

Current flowing through the surface of RF electrode generates an oscillating magnetic field $\bm{B}_1(t)= B_1 \sin(\Omega_{\text{RF}} t) \hat{\bm{y}}$, in addition to the stable magnetic field $\bm{B}_0$ from mounted permanent magnets $45\degree$ along the trap axis. Assuming $B_0 \gg B_1$, in a new coordinate system with $\x'$ aligned with $\bm{B}_0$ shown in Fig.~\ref{fig:exp_config}, we have the total magnetic field and the corresponding polarization vectors 
\begin{eqnarray*}
\bm{B}(t) &\approx& B_0 \left(\cos{a(t)} \x'+ \sin{a(t)}\y'\right) \\
\hat{\bm{\pi}}(t) &=& \cos{a(t)} \x'+ \sin{a(t)}\y'\\
\hat{\bm{\sigma}}_{\pm}(t) &=& \frac{1}{\sqrt{2}}\left( -\sin{a(t)} \x'+ \cos{a(t)}\y' \pm i\z'\right)
\end{eqnarray*}
with $a(t) = a \sin(\Omega_{\text{RF}} t) = B_1/(\sqrt{2}B_0) \sin(\Omega_{\text{RF}} t)$.

For a pure $\sigmap$ beam along $\bm{B}_0$ with $\bm{E}=E_0/\sqrt{2}(\y'+ i\z')$ and power $P_0$, its projections on the polarization components defined with respect to the direction of $\bm{B}$ are
\begin{eqnarray*}
\bm{E}\cdot\hat{\bm{\pi}}(t) &=& \frac{E_0}{\sqrt{2}}\sin{a(t)}\approx \frac{E_0}{\sqrt{2}}a(t)\\
\bm{E}\cdot\sigmap^*(t) &=& \frac{E_0}{2}(\cos{a(t)} +1)\approx E_0\left(1-\frac{a(t)^2}{4}\right)\\
\bm{E}\cdot\sigmam^*(t) &=& \frac{E_0}{2}(\cos{a(t)} -1)\approx -E_0\frac{a(t)^2}{4}.
\end{eqnarray*}
with averaged power 
\begin{eqnarray*}
P_{\hat{\bm{\pi}}} &\propto& \overline{(\bm{E}\cdot\hat{\bm{\pi}}(t))^2} =\frac{P_0}{2} \overline{a(t)^2} = P_0 \frac{a^2}{4}\\
P_{\sigmap} &\propto& \overline{(\bm{E}\cdot\sigmap^*(t))^2} \approx P_0\left(1-\frac{\overline{a(t)^2}}{2}\right) = P_0\left(1-\frac{a^2}{4}\right)\\
P_{\sigmam} &\propto& \overline{(\bm{E}\cdot\sigmam^*(t))^2} = P_0 \times \mathcal{O}(a^4)\approx 0, 
\end{eqnarray*}
where the bars over quantities indicate a time average. From these expressions we find that an oscillating field amplitude of $B_1 = 1.94$~G would fully explain the $\epsilon_\mathrm{imp}$ observed above. 

We estimate $B_1$ due to the current $I_{\text{RF}}(t) = I_0 \sin( \Omega_{\text{RF}}t)$ flowing along the RF electrodes. Considering the boundary condition of zero current at the electrode's end, $I_0 = \frac{1}{2} \Omega_{\text{RF}} C_{\text{trap}} V_{\text{RF}}$ at the trap center, where $\Omega_{\text{RF}}=2\pi \times 48.4$ MHz, $C_{\text{trap}}$ is measured to be 13.5 pF, and the RF voltage amplitude is simulated to be $V_{\text{RF}} \sim 40$ V for 4.4 MHz radial modes, close to our operating condition. Since the electrode dimensions and ion height $h = 50$ \textmu m is much smaller than the wavelength associated with $\Omega_{\text{RF}}$ MHz, we calculate the quasistatic oscillating $B$-field amplitude as
\begin{eqnarray*}
    B_1 =\frac{\mu_0 I_0}{\pi w_{\text{RF}}}\left(\tan^{-1}\left(\frac{w_0+w_\text{RF}}{h}\right)-\tan^{-1}{\left(\frac{w_0}{h}\right)}\right). 
\end{eqnarray*}
For our RF electrode width $w_{\text{RF}} = 70$ \textmu m, and the distance from the RF electrode's inner edge to the center axis $w_0 = 27.5$ \textmu m, this yields $B_1 = 2.7$ G. As we measured the trap capacitance after wirebonding it to the carrier PCB with its associated parasitics, this analysis somewhat overestimates the RF current and hence $B_1$, but the approximate agreement with that inferred from the observed polarization impurity indicates that these field oscillations account for the majority of the polarization impurity observed. 

\section{Limitations to SW EIT cooling \label{App:EITmodeling}}

We carry out fully quantum master equation simulations using the same methods described in our previous work \cite{xing2025trapped}, considering the relevant 8-level internal atomic structure and quantized motion, and incorporate experimental parameters and measured nonidealities to understand limitations to the observed EIT cooling performance. 

\begin{figure}[t!]
    \centering
\includegraphics[width=0.48\textwidth]{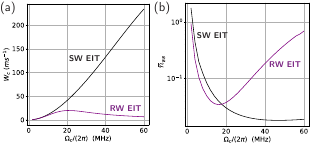}
    \caption{Simulated EIT cooling comparison incorporating known experimental limitations. (a) Cooling rate and (b) final steady-state phonon number for both SW and RW EIT cooling of $R_1$ as a function of cooling beam Rabi frequency $\Omega_c$.}
    \label{fig:cooling_rate}
\end{figure}

Fig.~\ref{fig:cooling_rate} shows the EIT cooling rate $W_c$ and final steady-state phonon number $\nbar_{\text{ss}}$ as a function of cooling beam's Rabi frequency $\Omega_c$ using the experimental detunings, pump beam amplitude, and motional frequencies. We also include the effects of all key nonidealities, namely the motional mode heating, pump beam polarization impurity, and residual $\Omega_c$ carrier coupling at the SW nodes. 

With $\Delta_p = \tpi120$ MHz, we take $\Delta_c = \Delta_p$, $\Omega_p = \sqrt{ (\Delta_p+2\omega_{R_1})^2-\Delta_p^2}$ for RW and $\Delta_c = \Delta_p + \omega_{R_1}$, $\Omega_p = \sqrt{8 \omega_{R_1} (\Delta_p+2\omega_{R_1})}$ for SW. We extract the cooling rate $W_c=-\dot\nbar/\nbar$ from the time evolution of $\nbar(t)$, starting from $\nbar(0) = 4$. While simulated RW EIT saturates at $W_c=20$ ms$^{-1}$ in line with the experimental measured $21(2)$ ms$^{-1}$, SW EIT's simulated $W_c$ increases further beyond the measured $57(3)$ ms$^{-1}$ with lower $\nbar_{\text{ss}}$ as $\Omega_c$ increases. This suggests that SW EIT can be significantly accelerated further by improving power delivery to the trap, and if positional drifts associated with photo-charging can be mitigated relative to those observed here (see Appendix~\ref{App:drift}). Matching the simulated cooling rates in this figure with those we measured, we estimate $\Omega_c = \tpi 21$ MHz for RW and $\Omega_c = \tpi 24$ MHz for SW; these are used to simulate the final phonon numbers achievable considering the various nonidealities in our experiment.

Table~\ref{tab:cooling_limit} shows the simulated final phonon number, achievable at the ideal and different unideal situations: heating rates, residual carrier excitation $\Omega_{\text{res}}$ from positional fluctuations or imperfect optical extinction, and $\sigmap$ beam impurity $\epsilon_{\text{imp}}$. This impurity is the primary limitation preventing cooling down to the $10^{-3}$ level, and can be addressed with RF electrode structures that minimize the oscillating magnetic field, or result in oscillating fields only along the direction of the static $B$-field.

We note that operating at lower $\Delta_p$ can reduce optical power required, as the pump beam intensity required for an ideal 3-level system is proportional to $\Omega_p^2 = 8\omega_m(\Delta_p + 2\omega_m)$ for motional frequency $\omega_m$, while the theoretical cooling rate and limit remain unaffected \cite{zhang2012dark}. To compare directly with the results in \cite{xing2025trapped}, we simulate SW EIT cooling at $\Delta_p = 2\Gamma$ and $B = 10$ G, in the ideal case for the 8-level $\Ca$ system, targeting $\nu = \tpi 3$ MHz. With the resulting $\Omega_p = \tpi34.2$ MHz and SW $\Omega_c = \tpi 42$ MHz, the total power $\propto \Omega_p^2 + \frac{\Omega_c^2}{2}$ is comparable to that used in recent work on polarization gradient cooling (PGC) in an integrated platform \cite{clements2024sub}. Starting from $\nbar = 4$, SW EIT cooling for these parameters reaches $\nbar_{\text{ss}} \approx 0.015$ within about 20 \textmu s. The higher $\nbar_{\text{ss}}$ compared to that achieved for $\Delta_p = 5\Gamma$ \cite{xing2025trapped} is due to greater deviation from the ideal 3-level system owing to larger off-resonant coupling to $\Pm$. However, the simulated $\nbar_{\text{ss}}$ and cooling time still suggest that SW EIT's advantage over PGC in terms of cooling limit holds at lower total powers comparable to those used for PGC.

\begin{table}[t!]
\begin{tabular}{cccccc}
       & Ideal               & Heating             & $\Omega_{\text{res}}$ & $\epsilon_{\text{imp}}$ & all errors \\ \hline
 $\nbar_{\text{ss}}^\mathrm{RW}$ & 0.03               & 0.038               &                       & 0.044                    & 0.052      \\ \hline
 $\nbar_{\text{ss}}^\mathrm{SW}$ & $6.8\times 10^{-5}$ & $2.8\times 10^{-3}$ & $1.6\times 10^{-3}$   & 0.021                   & 0.025     \\ \hline
\end{tabular}
    \caption{Simulated final phonon numbers $\nbar_{\text{ss}}^\mathrm{RW}$ and $\nbar_{\text{ss}}^\mathrm{SW}$ for RW and SW EIT cooling for the $R_1$ mode. We present these limits under ideal conditions with no technical noise or imperfection, and with contribution of individual nonidealities (heating, residual carrier amplitude at the SW node $\Omega_\mathrm{res}$, and pump beam polarization impurity $\epsilon_\mathrm{imp}$), and in the presence of all known nonidealities simultaneously. Simulations use experimental EIT parameters and estimated $\Omega_c$, and include effects of higher-order sidebands in the interaction Hamiltonian.}
    \label{tab:cooling_limit}
\end{table}

Table~\ref{tab:cooling_limit} presents the expected $\bar n_\mathrm{ss}$ limits under our experimental parameters, both in the absence and presence of the known nonidealities. We note that the ``ideal" limit predicted here is in fact limited by the $\Omega_c$ used in the experiment to achieve high cooling rates, which is only roughly 2$\times$ lower than $\Omega_p$ and thus perturbs the ideal EIT lineshape. The fundamental limit on phonon number of this scheme is achieved under the low intensity limit; near $\Omega_c = \tpi 5$ MHz and in the absence of technical noise, we simulate that SW EIT results in a steady-state phonon number of $\nbar_{\text{ss}} = 1 \times 10^{-5}$ for our experimental motional frequency and quantizing field $B_0 = 8.9$~G. This result is insensitive to the quantizing $B$-field strength as well as whether higher-order interaction terms are included in the simulation, and is in fact limited by sideband couplings in the RW pump beam used our experiments \cite{xing2025trapped}. 

While our model in the presence of known nonidealities explains the order of magnitude of $\bar n_\mathrm{ss}$ observed experimentally, it still predicts $\bar n_\mathrm{ss}$ approximately 2$\times$ lower. We expect that this additional factor of two discrepancy may be due to additional intensity-dependent positional deviations during EIT cooling sequences as compared to the AC stark shift measurements used to estimate these displacements, as well as uncertainties in our polarization impurity measurement. We have additionally considered the effect of relative phase fluctuations between the pump and cooling beams; modeled as Markovian noise, we find that relative linewidths of order 1 MHz would be required to explain this factor of two in final $\bar n_\mathrm{ss}$. This is far in excess of expected levels from acoustic noise and vibrations on the differential fiber paths these beams take prior to the device, but effect of phase fluctuations between the two driving tones may be relevant in experiments reaching significantly lower phonon numbers. Future experiments with suppressed quantization axis oscillations due to trap RF, as well as reduced technical noise, will ultimately be required to validate the significantly lower limits predicted by our model. 

We note that when implemented at practical cooling rates (and therefore effective excited state linewidth), even resolved sideband cooling is limited by off-resonant carrier and blue-sideband excitation \cite{leibfried2003quantum} that are entirely nulled in the SW EIT scheme. Our model thus indicates that if our present technical limitations can be addressed, SW EIT may enable even lower occupancies than achievable with resolved sideband cooling.

\section{Motional heating; light-induced drift \label{App:drift}}

\begin{figure}[b!]
    \centering
\includegraphics[width=0.48\textwidth]{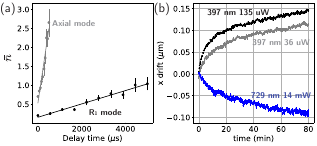}
    \caption{Electric field noise and drift characterization. (a) Phonon numbers of the axial and $R_1$ mode as a function of additional delay time after RW EIT cooling of the target mode. We measure the heating rate $\dot{\nbar}$ of axial and $R_1$ mode to be 3.9(3) ms$^{-1}$ and 0.18(1) ms$^{-1}$. (b) Trap position drift along $x$ due to integrated-laser-induced charging. Laser wavelength and input power before the feed-through are labeled in the plot. The drift at each time is inferred from the $x$-compensation field required to position the ion at a SW node. The three scans were taken on different days spanning over a week; in the multiple hours between scans during with the SW light was off, the trap position relaxed to the same initial value as shown. Error bars in both plots indicate 1$\sigma$ uncertainty.}
    \label{fig:noise_drift}
\end{figure}

We measure heating rates of secular motional modes of the trapped ion to infer electric field noise at the corresponding frequencies. Fig.~\ref{fig:noise_drift}a shows measured excitations $\bar n$ in both the axial and $R_1$ modes as a function of wait time after ground-state cooling, indicating heating rates of 3.9(3) ms$^{-1}$ and 0.18(1) ms$^{-1}$ respectively. The axial heating rate is consistent with that observed in the previous generation of fabrication \cite{mehta2020integrated}. 

We also characterized slow drifts in ion positioning due to laser-induced charging \cite{Wang_Hao_2011}. We quantify charging by measuring the ion's position drifts relative to the SW, using the AC Stark shift sequence. The ion's position shift is inferred from the change in applied DC field along $x$ required to position the ion at the SW node. As shown in Fig.~\ref{fig:noise_drift}b, the integrated 397 nm and 729 nm beams shift the ion in different direction along $\x$, possibly due to couplers' placement. The extent of drift in response to 397 nm light shows a modest power-dependence, while the comparable response to the 729 nm light for drastically higher powers suggests much stronger response to the UV photons. In both cases, upon starting the experimental sequences the ion experiences a rapid drift of 10s of nm that evolves into a slower constant drift. Between scans and over multiple hours, the trap position relaxed to nearly the same initial value, giving the same $x$ compensation at the target SW node at the beginning of each scan in Fig.~\ref{fig:noise_drift}, suggesting a reversible and reproducible laser-induced charging process. The observed drifts are comparable to those reported against a phase-stable polarization gradient in \cite{corsetti2024integrated}. In our experiments, data was taken generally in the limit where pulse sequences had been running for multiple minutes, such that ion positioning was stable to within $\sim10$ nm. For profiling and cooling experiments, we operated with up to 400~\textmu W incident in fiber before the feedthrough supplying the $\lambda = 397$ nm SW path. 

The slow trap location drifts particularly in response to UV light in the waveguide channel may suggest imperfect shielding by the ITO layer. The present 20~nm-thick ITO films exhibited sheet resistances of approximately 400 $\Omega$/square at room temperature, significantly higher than those of optimized films at this thickness. Influence of ITO film quality or surface preparation on the observed drifts would be an interesting and important subject for future study, as well as the relative performance of ITO vs. other transparent conductors in this regard \cite{wang2025can}.

\bibliography{coolingbib}

\end{document}